\RequirePackage{fix-cm}
\documentclass[smallextended]{svjour3}       
\smartqed  

\usepackage[colorinlistoftodos]{todonotes}
\usepackage{graphicx}
\usepackage{natbib}
\usepackage{color}

\usepackage{todonotes}
\usepackage{mathptmx}      
\usepackage{latexsym}
\usepackage{amssymb}
\def\lsim{\;\raise0.3ex\hbox{$<$\kern-0.75em\raise-1.1ex\hbox{$\sim$}}\;}
\def\gsim{\;\raise0.3ex\hbox{$>$\kern-0.75em\raise-1.1ex\hbox{$\sim$}}\;}

\newcommand{\apss}{Astrophys. Space Sci.  }

\newcommand{\apj}{ApJ }

\newcommand{\apjl}{ApJL }
\newcommand{\mnras}{MNRAS }
\newcommand{\aap}{A \& A }

\newcommand{\ssr}{Space Science Reviews }
\newcommand{\na}{New Astronomy }

\newcommand{\araa}{ARAA }

\newcommand{\physrep} {Physics Reports }
\newcommand{\memsai}{Memorie della Societ\'{a} Astronomica Italiana }

\newcommand{\jgr}{Journal of Geophysical Research 1}
\newcommand{\jcap}{Journal of Cosmology and Astroparticle Physics }
\newcommand{\aapr}{The Astronomy and Astrophysics Review }
\newcommand{\grl}{Geophysical Research Letters }

\newcommand{\enzo}{{{\small ENZO }}}

\newcommand{\pre}{Physical Review E }
\begin{document}

\title{Shocks and non-thermal particles in clusters of galaxies 
}


\author{A.M.Bykov, F.Vazza, J.A.Kropotina, K.P.Levenfish, F.B.S.Paerels}


\institute{
            \at
              second address \\
}

\institute{A.M.~Bykov \at
           Ioffe Institute, 194021, St. Petersburg, Russia;\\ 
           \email{byk@astro.ioffe.ru}  \and  F.~Vazza \at Dipartimento di Fisica e Astronomia, Universit di Bologna, Via Gobetti 93/2, Italy;\\Hamburger Sternwarte, Universit\"{a}t Hamburg, Gojenbergsweg 116, 40129 Hamburg, Germany \and J.A.~Kropotina \at
           Ioffe Institute, 194021, St. Petersburg, Russia\\
           \and K.P.~Levenfish \at
           Ioffe Institute, 194021, St. Petersburg, Russia\\
          \and  F.B.S.~Paerels \at Columbia Astrophysics Laboratory
Columbia University
550 West 120th Street, New York NY 10027
USA\\\email{frits@astro.columbia.edu}}

\date{Received: date / Accepted: date}

\maketitle
\begin{abstract}
Galaxy clusters grow by gas accretion, mostly from mergers of substructures, which release powerful shock waves into cosmic plasmas and convert a fraction of kinetic energy into thermal energy, amplification of magnetic fields and into the acceleration of energetic particles. The modeling of the radio signature of cosmic shocks, combined with the lack of detected $\gamma$-rays from cosmic ray (CR) protons, poses challenges to our understanding of how cosmic rays get accelerated and stored in the intracluster medium. Here we review the injection of CRs by cosmic shocks of different strengths, combining the detailed "microscopic" view of collisionless processes governing  the creation of non-thermal distributions of electrons and protons in cluster shocks (based on analytic theory and particle-in-cell simulations), with the "macroscopic" view of the large-scale distribution of cosmic rays, suggested by modern cosmological simulations. Time dependent non-linear kinetic models of particle acceleration by multiple internal shocks with large scale compressible motions of plasma with soft CR spectra containing a noticable energy density in the super-thermal protons of energies below a few GeV which is difficult to constrain by Fermi observations are discussed. We consider the effect of  plasma composition on CR injection and super-thermal particle population in the hot intracluster matter which can be constrained by fine high resolution X-ray spectroscopy of Fe ions.
\end{abstract}

\section{Plasma shocks  in structure formation} 
\label{sect.intro}
Accurate estimations of the energy content in the non-thermal components (cosmic rays and magnetic fields, beside turbulent gas motions) in clusters of galaxies are needed to allow for "precision cosmology" using galaxy clusters (see Pratt et al this topical collection). 
Direct evidence of the presence of the non-thermal components came from the radio observations of 
the extended to Mpc-size synchrotron emission in clusters of galaxies  \citep[see e.g.][]{1980ARA&A..18..165M,2008SSRv..134...93F,2010Sci...330..347V,2012A&ARv..20...54F,bj14},
while gamma-ray observations have so far only produced upper limits on the level of diffuse hadronic emission from the intracluster medium (ICM).
Hence non detections of hadronic $\gamma$-rays  from cosmic rays in clusters have in principle the potential not only to limit the total budget of  cosmic rays in clusters  \citep[][]{re03,gb07,ack10,fermi13,fermi14,2014ApJ...795L..21G,2016ApJ...819..149A,2014A&A...567A..93P,2014MNRAS.440..663Z,
2015A&A...578A..32Z,2017AIPC.1792b0009B} {\footnote{It shall be noted that these constraints are complementary to the ones that can be derived from radio analysis \citep[][]{de00,pe04,br07,donn10,brown11,bl11b,2012MNRAS.426..956B}, in which however the limits are intertwined with the uncertainties on the intracluster magnetic fields.}}, but also to limit the acceleration efficiency by merger shocks \citep[][]{va14relics,va15relics}. These can constrain the magnetic fields in clusters as well as the parameter space of the models of (re)acceleration of relativistic protons and their secondaries by turbulence in clusters \citep[see e.g.][]{2017MNRAS.472.1506B}.

\begin{figure*}
	\centering
    \includegraphics[width=0.99\textwidth]{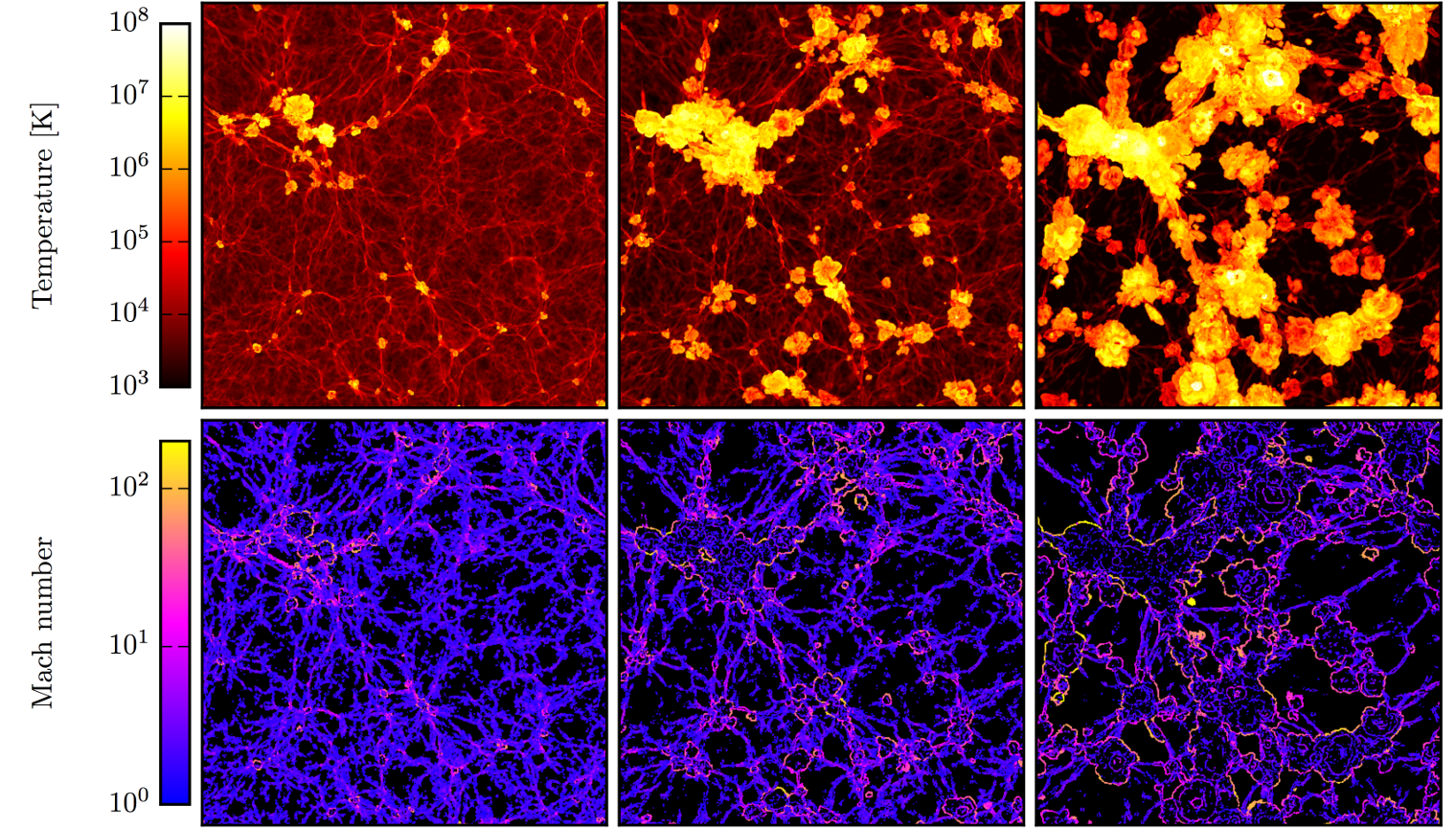}
    \includegraphics[width=0.49\textwidth,height=0.4\textwidth]{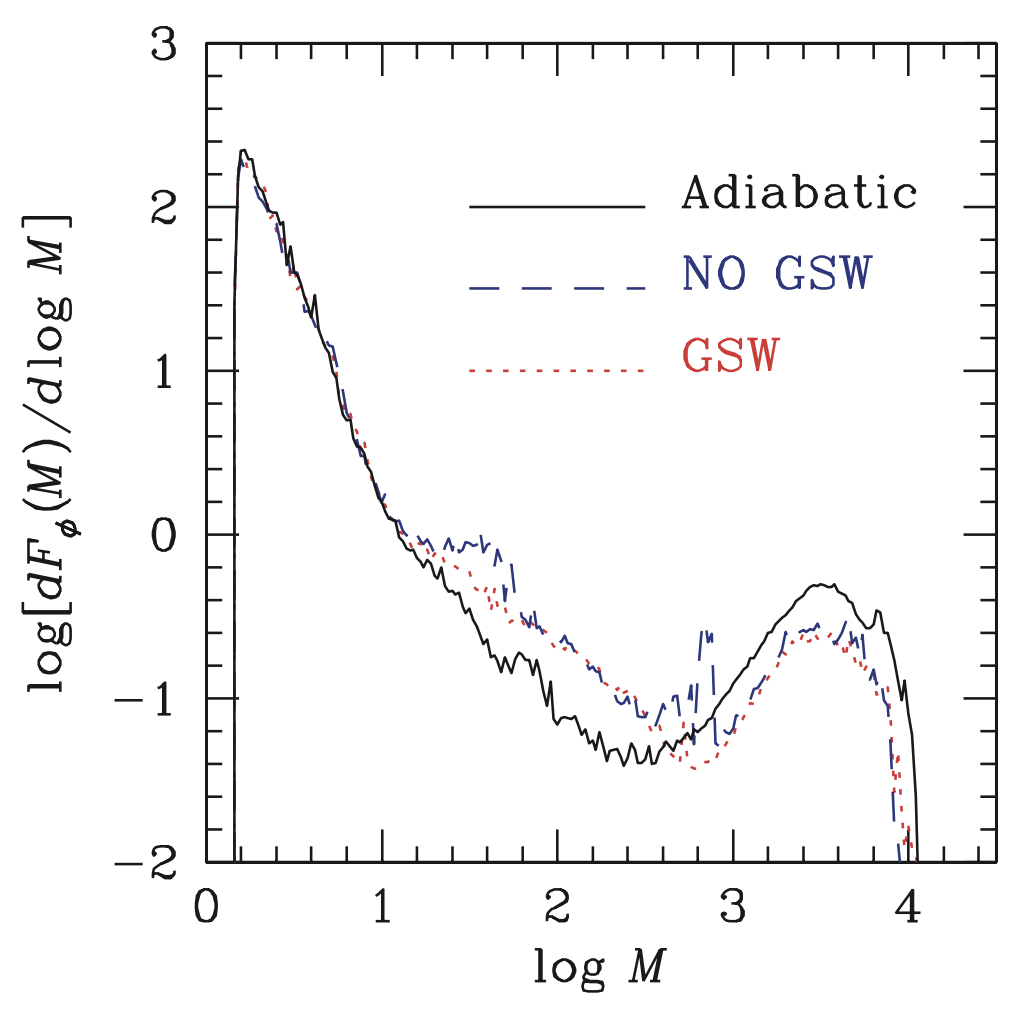}
    \includegraphics[width=0.49\textwidth,height=0.4\textwidth]{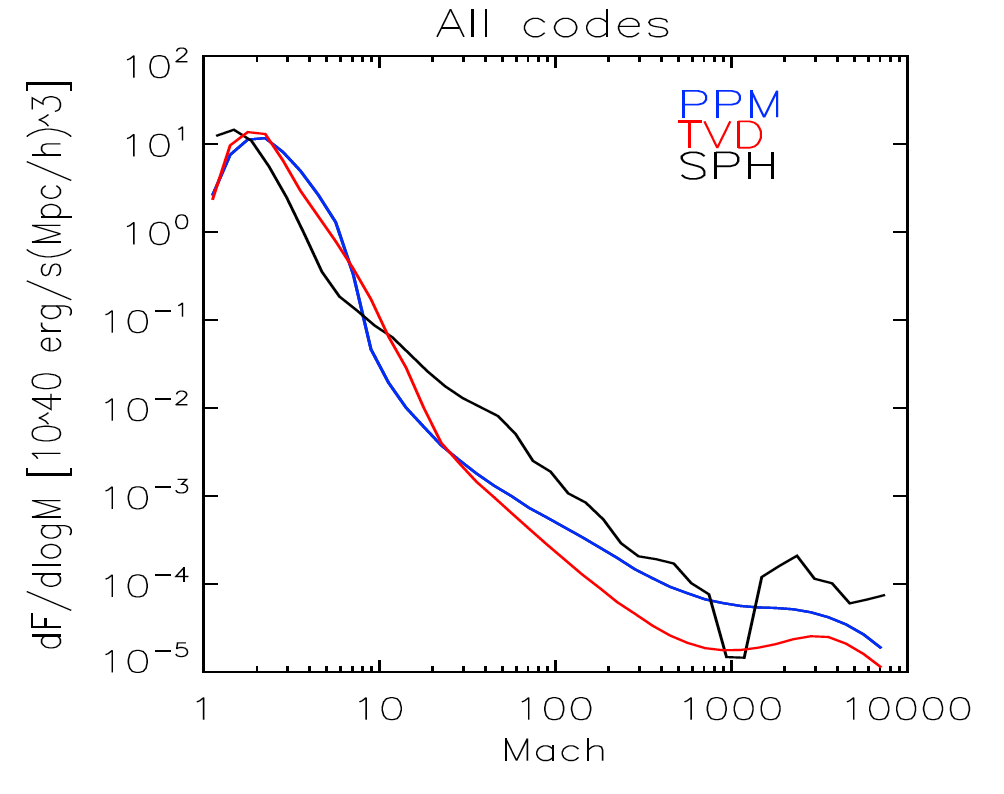}
    \caption{Spatial distribution and properties of shock waves on cosmological scales based on simulations. Top panels: evolution of gas temperature and shock Mach number for a thin slice crossing the Illustris-1 box at z=2, z=1 and z=0, from \citet{2016MNRAS.461.}. Bottom left panel: inverse of the mean comoving distance between shock surfaces at z=0 as a function of Mach number in a cosmic volume simulated by \citet{ka07}, comparing an "adiabatic" (i.e. non radiative model) with a radiative model with (GSW) or without (NO GSW) feedback from galactic star wind. Bottom right: distribution of dissipated kinetic energy flux as a function of shocks Mach number  identified in three resimulations of the same $(100 ~\rm Mpc/h)^3$ volume at z=0, employing  three different numerical methods: ENZO Eulerian simulations with the Piecewise Parabolic Method (PPM), Eulerian simulations with the Total variation diminishing (TVD) method \citep[][]{ry03}, and Smoothed Particle Hydrodynamics (SPH) simulations with GADGET \citep[][]{pf06}, adapted from \citet{va11comparison}} \label{fig:first}
\end{figure*}

The growth of cosmic structures cannot happen without the formation of shock waves, at whose surface kinetic energy gets dissipated into gas thermalization \citep[e.g.][]{1972A&A....20..189S,2007PhR...443....1M,by08,2009JCAP...08..002K,2012SSRv..166..187B,2013PhR...533...69C}. 
Around galaxy clusters, strong  $\mathcal{M} \geq 10-10^2$ accretion
shocks should exist in a quasi-stationary way and mark the transition from smooth infalling matter to 
halos undergoing their virialisation process, as illustrated in the top panels of Fig.\ref{fig:first}  \citep[][]{mi00,ry03,pf06,va09shocks,2013MNRAS.428.1643P,2016MNRAS.461.,2017Ap&SS.362...91M}. 
The statistics of shocks in the Universe is predicted to be dominated by structure formation shocks, with only a minor contribution from non-gravitational processes, as shown in the bottom left panel of Fig.\ref{fig:first} \citep[e.g.][]{ka12,va14curie,2016MNRAS.461.}.

Beside minor differences related to the different adopted hydrodynamical solvers and shock finders, cosmological simulations generally agree that   bulk of kinetic energy in the cosmic volume gets dissipated by $2 \leq \mathcal{M} \leq 4$ shocks within large-scale structures (see bottom right panel of Fig.\ref{fig:first} and \citealt[][for a comparison of cosmological codes]{va11comparison}). This range of  shock strength is intriguingly close to the typically one inferred  for shocks
associated with observed radio relics (see Van Weeren et al., this topical collection).  Such shocks are often associated with major mergers,  and these events 
are expected to maximally enrich the intracluster medium with cosmic rays. The typical regime of merger shocks is also the one at which making predictions of shock acceleration gets more difficult, owing to a number of physical and numerical uncertainties (see Sections below), hence the comparisons with real  $\gamma$-ray observations (or upper limits)  are particularly important to test  "micro"-details of shock acceleration models. \\

We notice that galaxy clusters represent an important testbed for any model attempting to describe the  acceleration of cosmic ray protons by structure formation shocks, in a plasma regime complementary to supernova remnants and solar wind shocks. This stems from the fact that, once accelerated, relativistic cosmic ray protons should be long-lived as they are subject to negligible energy losses, and must remain confined inside clusters for cosmological timescales  \citep[][]{1996SSRv...75..279V,bbp97,volk99}. $\gamma$-ray observations (or lack thereof) of the diffuse hadronic emission from the intracluster medium have in principle  the potential of testing theoretical scenarios of shock acceleration, because they can constrain the time-integrated budget of cosmic rays, injected over the entire lifetime of galaxy clusters.  

In addition to shocks, acceleration and re-acceleration of both electrons and nuclei by the compressible large scale plasma motions (where the energy containing scale of the turbulence is longer than the CR mean free paths) were both  proposed as an efficient Fermi type CR acceleration mechanism \citep[see][and the references therein]{1990ZhETF..98.1255B,1993PhyU...36.1020B,2000A&A...362..886B,2007MNRAS.378..245B,2011MNRAS.412..817B}.  
 Such a large scale turbulence can be induced by the merging and accretion of the cluster sub-structures.

The key problem of the cosmic ray modeling in clusters of galaxies is the transport of CRs,  which is governed by the magnetic fluctuations of very different scales which span about ten decades. Strong accretion shocks can convert about 10\% of their kinetic power to amplify magnetic fields and to built the extended spectra of magnetic fluctuations \citep[see e.g.][]{2014ApJ...789..137B} as it is likely the case in young supernova remnants. Precise microscopic modeling of the magnetic fluctuation spectra which are produced by both the power of the cluster formation and the different plasma instabilities during the cluster relaxation is non feasible now because it involves the very broad dynamical range of the fluctuations. Physics of the MHD waves in the hot cluster plasmas with high $\beta = 8 \pi n k_B (T_e + T_i)/B^2$ (where $n$ is particle number density, electron and ion temperatures are $T_e$ and $T_i$ and $B$ is magnetic field)  is a subject of the current interest \citep[][]{2018MNRAS.473.3095W,2018arXiv181007726X}.

We discuss the damping rates of the plasma long wavelength fluctuations to estimate the life time of the turbulence in section \ref{sec:LD}. Then in sections \ref{PICshock} --  \ref{S_ensemble} we consider CR injection and acceleration.   In section \ref{CS} we briefly discuss how cosmological simulations of evolving galaxy clusters have assisted the theoretical interpretation of $\gamma$-ray limits from galaxy clusters. 

\section{Plasma physics of the long-wavelength motions in clusters of galaxies}
\label{sec:LD}

The Coulomb mean free part of a thermal ion in a plasma of number density $n_{-3}$ (measured in the units of 10$^{-3}$ cm$^{-3}$ and temperature $T_8$ (measured in 10$^8$ K) can be estimated as $\lambda_C \approx 17\,  T_8^2/n_{-3}$ kpc.  Note that at the redshifts $z <0.05$, the superb  Chandra's  angular resolution corresponds to linear sizes of less than a kiloparsec. Therefore it is possible to study in X-rays different collisionless structures in the intercluster medium \citep[see e.g.][]{2007PhR...443....1M,2018NatAs...2..292W}.  
  
Compressible magnetosonic wave of a frequency $\omega$ propagating in high $\beta$ collisionless plasma is a subject of a strong Landau damping.  In an isothermal  maxwellian plasma with $T_e = T_i$ a fast magnetosonic wave of a phase velocity $v_{\rm ph}$ propagating at an angle $\theta$ to the mean homogeneous magnetic field of a magnitude $B_0$  would have a linear damping rate $\gamma$:

\begin{equation}
\frac{\gamma_{\rm ms}}{\omega} =
\sqrt{ {{\pi}\over 4} }
{{ v_{\rm Ti}}\over{v_{\rm ph}}}{{ \sin^2\theta}\over{|\cos\theta|}}
\Big [\sqrt{\frac{m_e}{m_p}} + 5 \exp \Big (- \frac{v_{\rm Ti}^2}{v_{\rm ph}^2 \cos^2\theta} \Big)
\Big ]. 
\label{Landau}
\end{equation}

Therefore in high $\beta$ {\sl Maxwellian} plasma where  the ion thermal velocity $v_{\rm Ti} >> v_{\rm ph}$ only nearly parallel propagating magnetosonic waves of $\theta <<1$ can  survive according to the linear approximation given by Eq.~\ref{Landau}. 
However,  the nonlinear effects of particle trapping, particle energy change due to scattering and trapping can result 
in a deformation of the initially maxwellian distribution function around the Landau resonant momenta \citep[][]{1965PhFl....8.2255O,1985crim.book.....T,1977PhFl...20.2093M,2010JMP....51a5204M,2018CNSNS..65..111A}. Some fraction of the particles can be  captured between crests of the wave and these particles do not change the energy of the wave.  Only uncaptured (transit) particles at the Landau resonant region efficiently take off the energy of the wave. In this regime the wave damping is determined by the collision rate $\nu_{\rm coll}$  which regulates the exchange between the captured and uncaptured particle populations \citep[][]{1985crim.book.....T}. The  rate $\nu_{\rm coll}$ can be either the Coulomb collision rate (which is rather low in the ICM) or the particle scattering rate by collisionless magnetic fuctuations. The linear regime of strong Landau damping will occure if the effective collisions are capable of restoring the Maxwellian particle distribution over the time period of the particle oscillation. Otherwise,  the wave damping may be strongly reduced comparing to the linear damping rate given by Eq.~\ref{Landau} after a possible formation of a flat  plateau region at the distribution function. Estimations of the damping rate at the nonlinear stage of the  Landau mechanism  $\gamma^{\rm nl}$ of a magnetosonic wave with the magnetic field amplitude $\delta B$ can be expressed as:
\begin{equation}
\frac{\gamma^{\rm nl}_{\rm ms}}{\omega} =
{{\nu_{\rm coll}}\over{\omega}} \Big({{B_0}\over{\delta B}}\Big)^{3/2} {{ \sin^{1/2}\theta}\over{\cos^2\theta}}
\Big [\sqrt{\frac{2m_e}{m_p}} + 5 \exp \Big (- \frac{v_{\rm Ti}^2}{v_{\rm ph}^2 \cos^2\theta} \Big)
\Big]
\label{NL_Landau}
\end{equation} 

 The effective  particle collision rate $\nu_{\rm coll}$ introduced into  Eq.~\ref{NL_Landau} in a phemonological way is determined by the magnetic fluctuations at the background ion gyro-scale  ($r_{\rm gi} \sim 10^{10}$ cm). In particular,  the firehose and mirror instabilities at
saturation in the marginal stability regime may provide $\nu_{\rm coll} \sim \beta S_l$, where $S_l$ is the linear shear magnitude 
\citep[][]{2014PhRvL.112t5003K}. 

Simulations by \citet[][]{1977PhFl...20.2093M} of nonlinear wave-particle interactions in a single sinusoidal magnetosonic mode of a finite amplitude demonstrated the test particle trapping, saturation, oscillation of the kinetic energy, and the diffusion of particles in velocity space. He has found that in a system without the Coulomb collisions the damping rate of the single magnetosonic wave is roughly  $\propto \delta B^{-3/2}$ which is consistent with Eq.~\ref{NL_Landau}.

A similar approach allows to estimate the Landau damping rate of Alfven mode in the non-linear regime of the distribution function flattening  as 
\begin{equation}
\gamma^{\rm nl}_{\rm a} \sim \gamma^{\rm nl}_{\rm ms} \times \Big ({{\omega}\over {\omega_ B}}\Big)^{1/2},
\label{NL_Landau_AW}
\end{equation} 
where $\omega_ B$ is the gyrofrequency of a background ion, and  for MHD waves the factor $\omega/\omega_ B < 1$. The difference between $\gamma^{\rm nl}_{\rm a}$ and  $\gamma^{\rm nl}_{\rm ms}$   is due to the Alfven wave polarization. 

In the case of the Alfven waves of relatively large amplitudes ${{\delta B}\over{B_0}} > {{\omega}\over {\omega_ B}}$ the ion reflections on the magnetic field maxima of the wave provided the mode damping rate of  

\begin{equation}
\frac{\gamma^{\rm nl}_{\rm a}}{\omega} \approx
{{1}\over {8 \pi^{3/2}}} \Big ({{\delta B}\over{B_0}}\Big)^{2} {{v_{\rm Ti}}\over{v_{\rm a}}}
\Big [\sqrt{\frac{m_e}{m_p}} +  \exp \Big (- \frac{v_{\rm Ti}^2}{v_{\rm a}^2} \Big)
\Big ].
\label{NL_AW}
\end{equation} 
which may dominate over the non-linear Landau damping given by Eq.~\ref{NL_Landau_AW}. At the same time the particle energization due to reflections contribute to the superthermal particles pool production.  
We shall discuss below in \S~\ref{PICshock} the production of the superthermal particles by reflections at the shock waves. 

The semi-qualitative analytic estimations discussed above were made for a simplified case 
of a single wave with well defined polarization, while in the ICM case we expect turbulent states with a broad range of interconnected scales. It means that the exact kinetic modeling should be performed with the selfconsistent particle-in-cell type simulations, which would allow to study the possible and likely departures of particle distribution function of the background plasma from the equilibrium Maxwellian distributions. This is very important for the accurate estimations of the non-thermal part of the plasma pressure in the clusters. 

One should note that in the low $\beta$ plasmas of the warm and cold phases  of the interstellar matter in galaxies  the particles which are in Landau resonance with MHD waves have momenta at the declining part of the distribution function. Hence it is energetically feasible to build up the plateau in this regime to avoid  the strong Landau damping. On the other hand it is not easy to build up the plateau in the hot interstellar plasma produced by supernovae  \citep[][]{1983ICRC....9..247B},  where multiple weak shocks may dominate the turbulence \citep[][]{1987Ap&SS.138..341B}. In the high $\beta$  plasma of a hot cluster it would be necessary to build up the plateau close to the peak of the distribution, which is energetically very demanding. If the departures from the Maxwellian ion distribution are  present indeed,  they can be constrained by the fine ion spectroscopy of the hot X-ray emitting plasma. Most likely these highly turbulent regions are not spread over the whole relatively quiet ICM but are  intermittent and localized  in the vicinity of the fast moving interacting structures. 

Therefore in the cluster environment the collisionless compressible and incompressible modes of wavelengths $\lambda < \lambda_C$  are likely produced by fast motions of the relaxing cluster sub-structures, plasma instabilities and turbulent cascading of the large scale motions associated with shocks. The magnetic fluctuations of scales $\lambda \sim 10^{12} - 10^{17}$ cm  which   are responsible for resonant scattering of GeV-TeV regime cosmic rays in the intercluster medium    have scale sizes well below  $\lambda_C$, but much larger than the thermal ion Larmor-scale. 

Hybrid-kinetic numerical simulations by \citet{2014PhRvL.112t5003K} revealed that  momentum and heat transport in plasma structures with a persistent shear can be governed by magnetic fluctuations of $\delta B/B \sim 1$ at the ion- and sub-ion-Larmor-scales produced by firehose and mirror instabilities in a hot collisionless plasma in clusters of galaxies \citep[see also][]{2008PhRvL.100h1301S,2015ApJ...798...90Z,2016MNRAS.460..467K}. The strong magnetic field fluctuations of the thermal ion Larmor-scale are important for the plasma heat conduction and viscosity, but they can not scatter efficiently CRs.

 The very extended dynamical range  of the fluctuation wavenumbers in the system makes the full scale particle-in-cell  simulations to be unfeasible by now. In the meantime particle-in-cell simulations of the structure of collisionless shocks in high $\beta$ plasma can be successfully done given that much  narrower dynamical range of scales are involved there. The simulations of collisionless shocks demonstrate clear departures from the Maxwellian distribution of ions which provide microscopic justification of the local effects of particle injection into Fermi acceleration on macroscopic scales of we shall discuss this in \S~\ref{PICshock}.

\section{Particle-in-cell  modeling of collisionless shocks in hot cluster plasmas}\label{PICshock}

Shocks of different scales are playing a crucial role at different stages of cluster formation and evolution transforming the energy of the gas bulk motions to the plasma heating, turbulence production and particle acceleration.
Collisionless shocks in clusters  can locally produce a population of superthermal particles some fraction of which may be further accelerated/re-accelerated to long-lived relativistic component.  Collisionless shock microphysics is essential at the stages of shock formation, shock dynamics, plasma heating  and particle acceleration. Therefore the collisionless shock phenomenon is a multi-scale strong coupling non-linear problem in time and space  \citep[e.g.][]{2016RPPh...79d6901M}. At the scales of some tens of the proton gyroradii particle-in-cell (PIC) simulations are the most efficient way to study the microphysical structure 
of the viscous velocity jump and plasma ion heating in high $\beta$ cluster plasmas \citep[see e.g.][]{2010ApJ...721..828K}.

Electron heating in the collisionless shocks is a widespread phenomena at all astrophysical scales from the bow shocks in the Earth and Saturn and the solar flares to  supernova remnants and the cluster of galaxies scale. The electron heating efficiency (e.g. the postshock electron to proton temperature ratio) is very different for the collisionless shocks of different Mach numbers in the various environments \citep[see e.g.][]{2015A&A...579A..13V}.          
The electron heating by shocks with Mach numbers $\mathcal{M} \sim 3$ (in the range suggested by observations of cluster radio shocks, see Van Weeren et al. this topical collection) in hot high $\beta$ plasma of clusters of galaxies were studied  with particle-in-cell  modeling by \citet[][]{2017ApJ...851..134G,2018ApJ...858...95G}.  They found that the electron whistler instability which can break  the electron adiabatic invariance may occur due to the proton cyclotron and mirror  
modes accompanying the relaxation of proton temperature anisotropy. Then the non-adiabatic effects provide  an efficient electron entropy production at the microscales before the Coulomb relaxation at much longer scales \citep[][]{2008SSRv..134..141B}.  

Electrons can be directly injected to the non-thermal energies at the collisionless shocks to provide the primary leptonic component to be further energized in clusters of galaxies. Particle-in-cell simulations of the shocks of the sonic Mach numbers $\mathcal{M} \sim 3$ in the hot cluster plasmas were performed by \citet[][]{guo14a,2014ApJ...797...47G}. They found that in quasi-perpendicular shocks about 15\% of the initially thermal particles can be accelerated by the shock drift acceleration to a non-thermal power-law distribution of a slope about 2.4. \citet{2015JPhCS.642a2017M} found that the non-thermal electron acceleration by quasi-transverse shock is not explained merely by the shock drift acceleration, but the local wave-
particle interactions can play an important role \citep[see also][]{2017ApJ...843..147M,2018ApJ...857...36Y}.  

Combined hybrid particle-in-cell and test-particle simulations were used by 
\citet[][]{2018MNRAS.tmp.2630T}  to model the influence of  shock surface fluctuations on the acceleration of suprathermal electrons in three dimensional shock structure and to compare the 2D and 3D cases.  The 3D structure of the shock front allowed a possibility for the electrons to interact with more than one surface fluctuation per interaction which may increase the electron energization. They also found that somewhat unrealistic electron trapping apparent in 2D simulations may cause the higher energization in the subcritical 2D shocks.

Contrary to the electron case the proton injection is efficient in quasi-parallel shocks and inefficient in the quasi-perpendicular geometry \citep[][]{2014ApJ...783...91C}. The microscopical process of proton injection in high beta plasmas can be studied by PIC simulations.  \citet{2018ApJ...864..105H} have recently simulated  the reflection of 
incident protons and the self- excitation of plasma waves by CR-driven instabilities in the collisionless shocks under typical ICM conditions. They reported that only in ICM shocks with $\mathcal{M} \geq  2.25$  a sufficient fraction of incoming cosmic rays which are reflected at the shock (through a combination of the overshoot at the shock electric potential and magnetic mirror) produce further excitation of magnetic waves via CR streaming, hence starting diffusive shock acceleration. The simulated abundance of the merger shocks in 
the clusters is decreasing with the Mach number  above $\mathcal{M} \geq  2.25$ \citep[see e.g.][]{ry03}.
Therefore, in this scenario only the relatively small fraction of merger shocks with $\mathcal{M} \geq 2.25$ might be able to inject cosmic rays in galaxy clusters, potentially alleviating the tension with $\gamma$-ray limits. The authors defined the injection fraction as the number of non-thermal ions with energies above 10 thermal proton energies in the shock downstream and found that the fraction ranges between $10^{-2} - 10^{-3}$ for quasi-parallel shocks with $2.25 \leq \mathcal{M} < 4$. The PIC simulations discussed above were performed for proton electron plasmas with the typical mass ratio $m_p/m_e = 100$. 

Apart from the radiation of the relativistic particles the non-thermal injection phenomena can be potentially constrained with the X ray spectroscopy \citep[][]{2009A&A...503..373K}. Therefore it is interesting to study the injection processes in multicomponent plasmas with account for at least some abundant ions in addition to protons and electrons. This would require the larger simulation domain and the longer run times than that in electron-proton plasmas. Hybrid simulations which  treat ions kinetically while the electrons as a fluid are providing a way to extend the domain with the given computer resources. The hybrid codes  \citep[see e.g.][]{1988JGR....93.9649Q,1996JGR...10117287W,1997JGR...10219789G,2002hmst.book.....L,2007CoPhC.176..419G,2014ApJ...783...91C,2016JTePh..61..517K}
are widely used for modeling of collisionless shocks. We discuss in \S~\ref{sec:Hybr} the effect of He ions on the shock injection process and the shape of He and Fe ion distributions at the collisionless shocks  in high $\beta$ ICM plasma.

\subsection{Hybrid simulations of shocks in hot multicomponent plasmas}\label{sec:Hybr}

To address the effect of a realistic cluster plasma composition on the acceleration of CRs and the resulting departures of the ion heating from that given by the Rankine-Hugoniot equations  we present in this section some results of the hybrid simulations of shocks in high $\beta$ multi-ion cluster plasmas.  The  simulations discussed below were performed with the kinetic, three dimensional second order accurate, divergence-conserving hybrid code \textit{Maximus} \citep[][]{2018arXiv180605926K}. 
Hybrid codes are a special class of kinetic codes which combine Vlasov-Maxwell equations for ions with hydrodynamic equations for electrons. The Vlasov equation for ions is solved within the Lagrangian approach, i.e. tracing the individual macro-particles in the continuous phase space. Each macro-particle represents the ensemble of real ones. The electrons are treated as a massless neutralizing fluid, which leads to the following equations system:
\begin {eqnarray}
	&&\frac{d \vec r_k}{dt}=\vec V_k  \\
	&&\frac{d \vec V_k}{dt}=\frac{Z_k e}{A_k m_p}\left(\vec E + \frac{1}{c}\vec V_k \times \vec B \right) \\
	&&\frac {\partial \vec B}{\partial t} = - c\bigtriangledown \times \vec E \\
	&& \vec E = \frac{1}{4\pi \rho_c} \left(\bigtriangledown \times \vec B \right) \times \vec B 
                  - \frac{1}{c \rho_c} \left( \vec j_{\rm ions} \times \vec B \right ) - \bigtriangledown P_e / \rho_c  \\
        && \vec j_{\rm ions} = \sum_{\rm cell} S(\vec r_k) Z_k e V_k\\ 
        &&\rho_c = \sum_{\rm cell} S(\vec r_k) Z_k e 
	\end {eqnarray}
where $\vec r_k, \vec V_k, Z_k, m_k$ represent coordinates, velocities, charge and mass of macro-particles, $\rho_c, \vec j_{ions}$ - ions charge density and current, $P_e$ - electrons pressure, and $S(\vec r_k)$ - quadratic weighting function for charge deposit. Hereafter coordinates are given in units of proton inertial lengths $l_i = c/\omega_p$, velocities in alfvenic ones, and the mass density and magnetic field are normalized to their far upstream values. For the number density of ions, $n$, in multispecies plasma of mass density $\rho$
we took the proton concentration in the pure hydrogen plasma of the same $\rho$ \citep[][]{1994JCoPh.112..102M}.
The equation system must be closed by the electrons equation of state, which is actually unknown. The quasineutrality condition yields $n_e = n_i \equiv n$, so only the electrons downstream temperature is left to be determined to obtain $P_e$. The latter was taken from the PIC simulations with the code Tristan-MP (developed by: A.Spitkovsky, L.Gargate, J.Park, L.Sironi on the base of original TRISTAN code by O. Buneman.), and appeared to be consistent with the analytic estimations from \cite[][]{2015A&A...579A..13V}, supposing 5\% energy exchange between electrons and ions, i.e. $T_e = T_i$ for $\mathcal{M}_s = 2.0$ and $T_e = 0.7T_i$ for $\mathcal{M}_s = 3.0$.
The resulting equation system was solved numerically on a 3d cartesian grid with 50 macro-particles per species  per cell (ppsc) in the far upstream, which resulted in more than 150 ppsc downstream, due to the shock compression. The simulation box spanned $50,000 \times 1 \times 200$~cells of size $(0.5 \times 0.5 \times 0.5)\, l_i\,$.

The shock was initialized via the reflection of hot super-sonic super-alfvenic flow
traveling in the negative $x$-direction from the conducting wall at the $x = 0$ plane (the \textit{piston method}). The upstream ratio of thermal to magnetic pressure $\beta$ was taken equal to 100, with equipartition between electrons and ions. The far upstream  magnetic field $B_0$ was placed in the $x$-$z$ plane and oriented obliquely to the flow, at a small angle $\theta = 13^o$ to the shock normal~$x$. In the simulation (i.e. downstream) frame the initial flow alfven Mach numbers were $\mathcal{M}_a = 10$ and $\mathcal{M}_a = 20$, which led to the sonic Mach numbers (in shock rest frame) equal to $\mathcal{M}_s = 2.0$ and $\mathcal{M}_s = 3.1$ respectively. Note  that a  substantial admixture of He would change the effective sonic Mach number.  

\begin{figure*}
	\centering
    \includegraphics[width=0.9\textwidth]{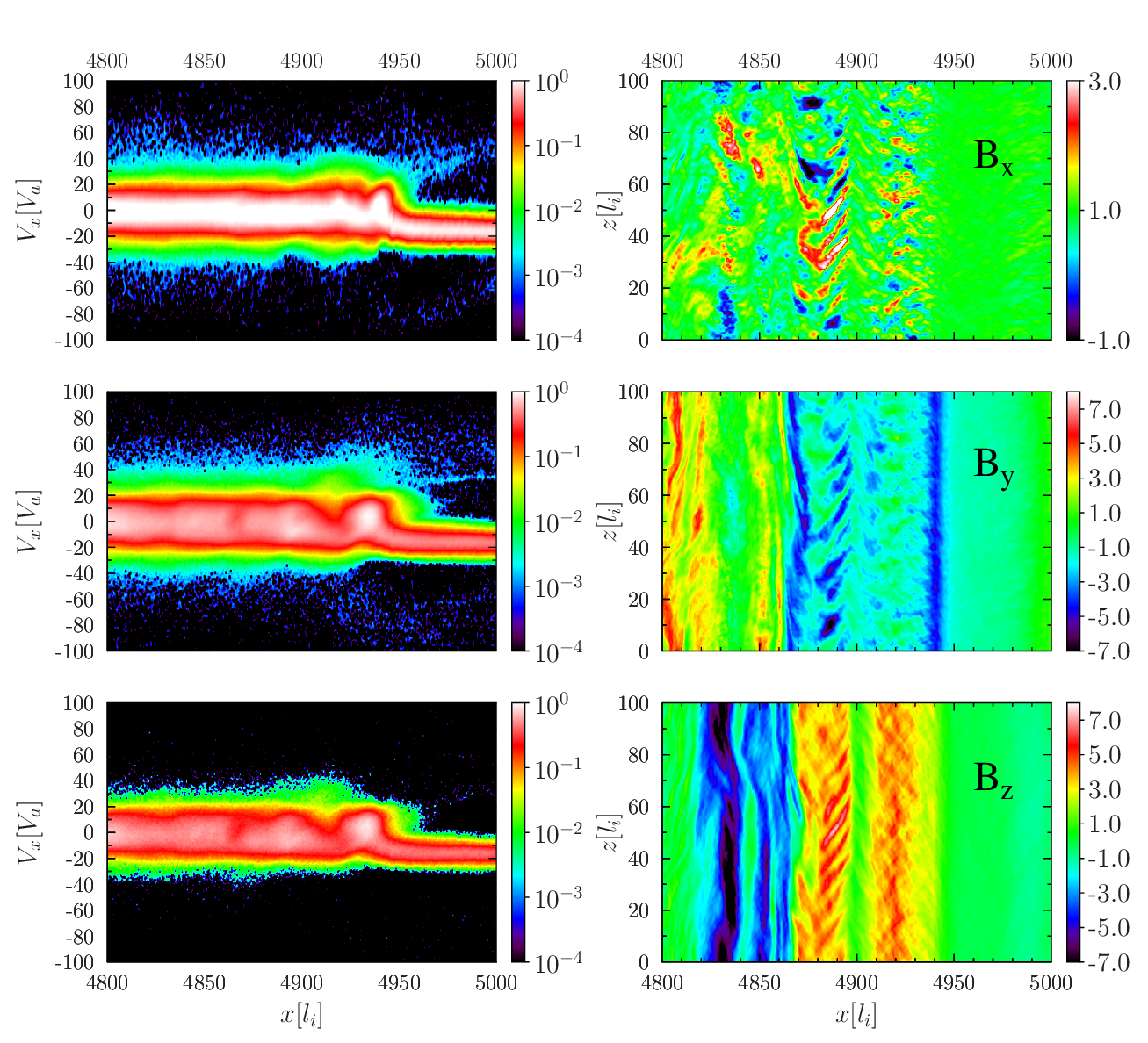}
    \caption{Simulated structure of the shock with the following upstream parameters: $\mathcal{M} \approx 3$ (in the shock rest frame), $T_e = T_i = 10^8 K$, $\beta = 100, \theta = 13^o$, solar abundance (75\% H II, 25\% He III, 0.1\% Fe XXVI by mass). The results are given at time $t = 600 \Omega^{-1}$ after shock initialization. The corresponding upstream alfven Mach number in the simulation frame is $\mathcal{M}_a = 20$. The $x-V_x$ phase spaces of protons, He$^{+2}$ and Fe$^{+25}$  are given in the left panel from top to bottom respectively, while the magnetic field components are color-coded in the right panel.} \label{fig:ShStructure}
\end{figure*}

The results of the hybrid simulations of the structure of the shock with the upstream parameters: $\mathcal{M} \approx 3$ (in the shock rest frame), $T_e = T_i = 10^8 K$, $\beta = 100, \theta = 13^o$, solar abundance (75\% H II, 25\% He III, 0.1\% Fe XXVI by mass)  are presented in Fig.~\ref{fig:ShStructure}. The structure is given at time moment $t = 600 \Omega^{-1}$ after shock initialization. 

\begin{figure*}
	\centering
    \includegraphics[width=0.9\textwidth]{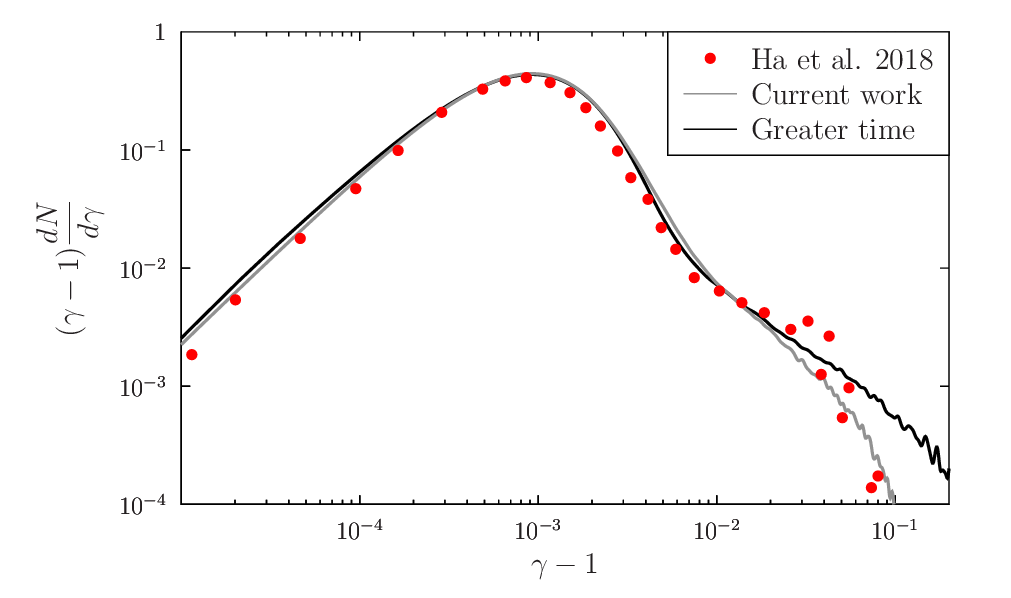}
    \caption{Far downstream proton energy spectra for the same shock as in Fig.~\ref{fig:ShStructure} at $t = 150\Omega^{-1}$ (gray curve) and $t=600\Omega^{-1}$ (black curve). The corresponding spectrum from \citet{2018ApJ...864..105H}  (run M3.2) at $t = 90\Omega^{-1}$ is shown by red dots. The consistency is within the spatial and temporal variations of the power law tail.} \label{fig:SpectraHa}
\end{figure*}

Then in Fig.~\ref{fig:SpectraHa} we compare the spectra of protons obtained in the long run hybrid simulations to the PIC simulations by \citet{2018ApJ...864..105H} at $t = 90\Omega^{-1}$. Far downstream proton energy spectra presented in this figure are for the same shock as in Fig.~\ref{fig:ShStructure} at $t = 150\Omega^{-1}$ (gray curve) and $t=600\Omega^{-1}$ (black curve). The corresponding spectrum of protons from \citet{2018ApJ...864..105H} (run M3.2) at $t = 90\Omega^{-1}$ is shown by red dots. The consistency is apparent within the spatial and temporal variations of the power law tail. 

\begin{figure*}
	\centering
    \includegraphics[width=0.9\textwidth]{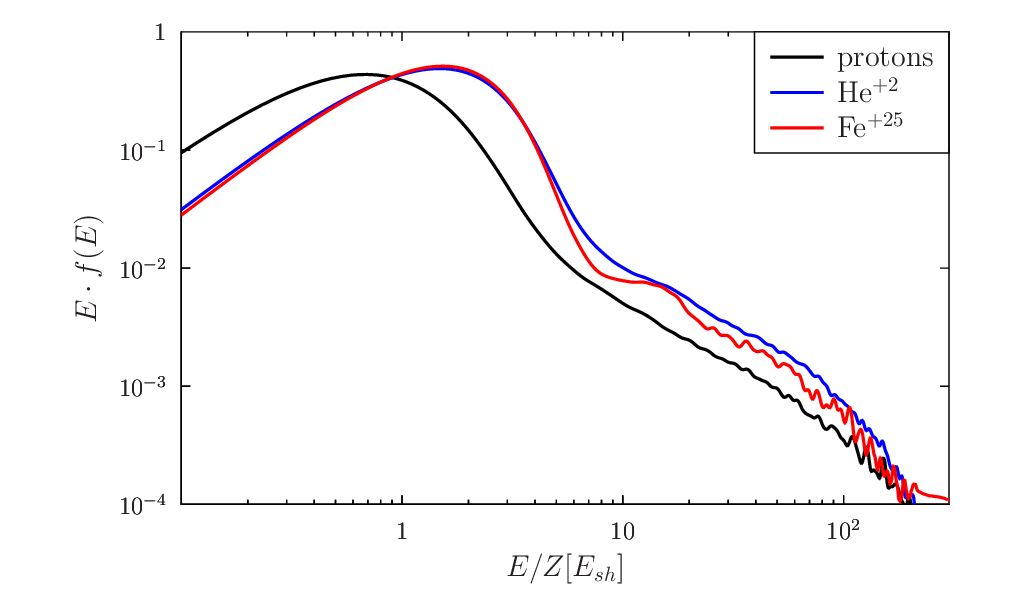}
    \caption{Far downstream energy distributions of protons (black), He$^{+2}$ (blue) and Fe$^{+25}$ (red) for the same shock as that in Fig.~\ref{fig:ShStructure} at $t = 600 \Omega^{-1}$. The relaxation of Fe is still in progress at this time. The distributions are normalized by unit, and the energy is given in terms of $E_{sh} = 0.5m_p V_{sh}^2 = 200 [m_p V_a^2]$ (here $V_{sh}$ is in the simulation (i.e. downstream rest) frame).} \label{fig:SpectraHE}
\end{figure*}

In Fig.~\ref{fig:SpectraHE} far downstream energy distributions of protons (black), He$^{+2}$ (blue) and Fe$^{+25}$ (red) are shown for the same shock as that in Fig.~\ref{fig:ShStructure}  at $t = 600 \Omega^{-1}$. The relaxation of the Fe ion is still in progress at this time. The distributions are normalized by unit, and the energy is given in terms of $E_{sh} = 0.5m_p V_{sh}^2 = 200 [m_p V_a^2]$ (here $V_{sh}$ is in the simulation (i.e. downstream rest) frame).

\begin{figure*}
	\centering
    \includegraphics[width=0.9\textwidth]{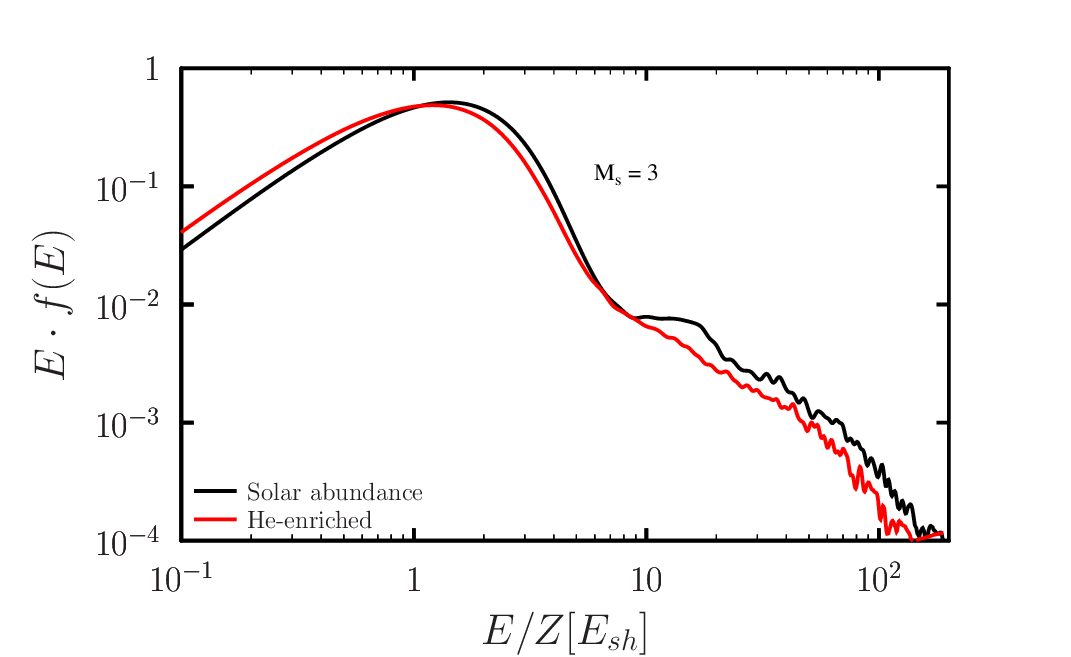}
    \caption{Far downstream energy distributions of Fe$^{+25}$  for different local abundances of He$^{+2}$ in the intercluster medium. The energy distributions of Fe$^{+25}$  for the solar (black) and for double (red) abundances  of He$^{+2}$ simulated  for the same shock  at $t = 600 \Omega^{-1}$. The energy is given in terms of $E_{sh} = 0.5m_p V_{sh}^2 = 200 [m_p V_a^2]$ (here $V_{sh}$ is in the downstream rest frame).} \label{fig:SpectraFE}
\end{figure*}

The multi-ion composition of the ICM may have a prominent effect on both the temperatures and non-thermal tails of the ions passed through the collisionless shock and hence can be constrained with the X-ray spectroscopy of galaxy clusters. Theoretical models predicted an important role of the diffusion and sedimentation effects in the redistribution of elements in the hot interstellar and intracluster media   \citep[see e.g.][]{1977MNRAS.181P...5F,1984SvAL...10..137G,2004MNRAS.349L..13C,2006MNRAS.369L..42E,2017AstL...43..285M,2018MNRAS.476.2689B}. X-ray observations of the galaxy clusters  \citep[see e.g.][]{2006A&A...452..397D,2010A&ARv..18..127B, 2017A&A...603A..80M} provided clear evidences  that the ICM does not have a primordial composition, but has been enriched with metals up to  about 1.5 - 2  times of the solar abundance. 

The helium abundance affects seriously the structure of electromagnetic fields in the shock transition region and the resulting ion distributions in the shock downstream. In Fig.~\ref{fig:SpectraFE} the far downstream energy distributions of Fe$^{+25}$ ion are shown for different possible abundances of He$^{+2}$ ions in the upstream flow. The solar abundance of He$^{+2}$ results in the energy distribution of Fe$^{+25}$ shown with the black curve,  while the red curve is the Fe$^{+25}$ distribultion in plasma with the doubled solar abundance of the helium.  The next generation of X-ray observatories \citep[e.g.][]{2018SPIE10699E..1GB}  may be used to constrain the helium abundance in clusters with the precise line spectroscopy of metals (e.g.  Fe$^{+25}$).

\subsection{The energy losses and the non-thermal tails}
While particle energy gain dominates over energy losses due to particle interactions in the vicinity of the shock of some thousands of gyroradii, at much larger distances the energy and momentum losses may become important.  

The rate of energy losses $dE_{\rm C}/dt$ of a test ion of velocity $v$ due to the Coulomb collisions in fully ionized plasma of the electron temperature $T_{\rm e}$ can be approximated as 
\citep[][]{2002cra..book.....S}

\begin{equation}
- \frac{dE_{\rm C}}{dt} = 3.1 \times 10^{-7} n_{\rm e} Z^2 \frac{(v/c)^2}{(x_m)^3 + (v/c)^3}~~ {\rm eV~ s^{-1}}, 
\label{Coulomb}
\end{equation}
where $x_m \approx 0.03 (T_{\rm e}/2 \times 10^6 K)^{1/2}$.

\begin{figure*}
	\centering
    \includegraphics[width=0.9\textwidth]{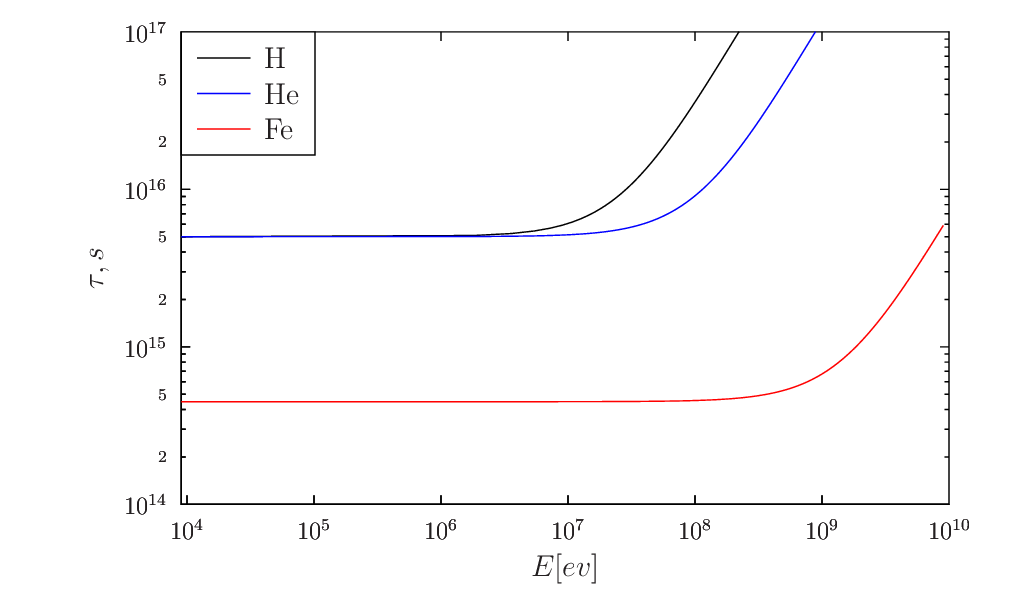}
    \caption{The lifetime of nuclei against the energy losses (the Coulomb losses in the fully ionized ICM) protons (black), He$^{+2}$ (blue) and Fe$^{+25}$ (red). The lifetime at the higher energies is limited by the inelastic nuclear interaction processes at about 10$^{17}$ s. } \label{fig:Losses}
\end{figure*}

At non-relativistic ion energies the lifetime of nuclei against the energy losses due to the inelastic Coulomb  collisions in the fully ionized hot ICM can be obtained by the integration of Eq.~\ref{Coulomb}. The lifetimes $\tau_{\rm s}$ for s = p,He$^{+2}$ and Fe$^{+25}$  are shown in Fig.~\ref{fig:Losses}. The inelastic nuclear collisions provide a fragmentation of He, Fe and the other nuclei even at the MeV energy regime with a characteristic lifetime longer than that is due to the Coulomb losses. Above the pion production threshold the energy losses are dominated by this process and the lifetime is limited to below 10$^{17}$ s in typical hot ICM conditions. The energy losses of electrons in the ICM where the synchrotron and inverse-Compton processes are important were discussed earlier in \citet[see e.g.][]{2008SSRv..134..207P} . 

An  important parameter is the width ($\Delta_{\rm s}$) of the region around the CR injector/accelerator (e.g. shock) filled with injected CRs. It can be estimated as $\Delta_{\rm s} \sim \sqrt{2 v ~\tau_{\rm s} ~\Lambda}$ (where $\Lambda$ is the particle mean free path) assuming the diffusive propagation of CRs in the cluster.    
If $\Delta_{\rm s}$ is larger than the mean distance between the injectors than apart from the individual shocks  of $\mathcal{M} \geq 2.25$ also large scale plasma motions and the weaker shocks can re-accelerate the seed population of the super-thermal particles. For injected particles of energy above 10 E$_{\rm sh}$ the mean free path due to the elastic Coulomb scatterings on the fully ionized ICM is above a Mpc, which mean that they would spread over the most of the ICM. However, if a magnetic turbulence of scales well below a parsec has a sizeable filling factor in the ICM, then the mean free path $\Lambda$ will be much smaller there and the Coulomb losses would be a factor. 
On the other hand the magnetic turbulence providing efficient scattering of non-thermal particles would allow CR re-acceleration by the long-wavelength compressible plasma motions of sub-Mpc scales  produced by the merging/accretion processes. We shall discuss the processes now.           

Apart from particle injection and acceleration by the individual shocks  the large scale plasma motions and weaker shocks may re-accelerate particles from the long-lived pool of non-thermal particles. In \S~ \ref{NLensemble} we discuss  the  spectra of the non-thermal particles re-accelerated by the multiple shocks and the associated compressible plasma motions.  

\section{Particle acceleration by ensemble of shocks in clusters of galaxies}
\label{S_ensemble}
The accretion and merging processes, as well as fast supersonic motions of dark matter halos with baryonic component should launch shocks in the ICM, from a megaparsec scale down to about 10 kpcs,  where Coulomb collisions are still important. The kinetic energy of large scale motions is a huge reservoir of power of the acceleration of particles and magnetic field amplification.  However, particle acceleration by the Fermi mechanism, which allows to transfer a part of the energy of the plasma motions to relativistic particles, can only proceed in presence of  effective particle scatterings by magnetic fluctuations.  This can be provided in the presence of magnetic fluctuations of scales comparable to the gyroradii of the accelerated particles. In the typical ICM magnetic fields($\sim ~\rm \mu G$) the scales of the resonant fluctuations are below 10$^{17}$  cm, i.e. well below the energy containing scale of the plasma motions.  

If the magnetic fluctuations of scales $l$ in the range  $10^{12} - 10^{17}$  cm exist in the collisionless regime and are a part of the extended Kolmogorov type  turbulent spectrum  of spectral index $\nu$  with the energy containing scale $L$, then 
the mean free path of CR particles can be estimated from the quasi-linear theory as $\Lambda(p) \sim g(\nu)  L (r_g/L)^{2 - \nu}$ where 
$g(\nu)$ is  a numerical factor \citep[see e.g.][]{1985crim.book.....T}. Then the parameter  $\eta(p)$ can be expressed as
\begin{equation}
\eta(p)  = uL/v\Lambda(p) = u/c \times  g^{-1}(\nu)  (L/r_g)^{2 - \nu}.
\end{equation}

For a typical bulk motions of the velocity amplitude  $u \gsim 1,000$ km/s  at $\geq$ 100 kpcs scales, and with the extended Kolmogorov type magnetic  turbulence,  one has $\eta(p) >$1.  For cosmic rays of low enough energies to satisfy $\eta(p) >$1,  the CR distribution function $F({\mbox{\boldmath $r$}},p,t)$ is highly intermittent in space,  with sharp concentration gradients in the vicinity of the shocks and the smooth backround distribution in between of the shocks. Contrary to the case of the low energy CRs  with  $\eta(p) >$1,  in the high energy regime the typical distance between the shocks is $L <l_{\rm sh}(p)  = c/u \times \Lambda(p)$ which is characterizing the width the distribution of CR of momentum $p$ in the shock vicinity. Therefore the condition of  $\eta(p) < $1 provides a smooth CR distribution at the high energy end of the accelerated particles.  

Since the particle distribution within ICM  with the shock ensemble is highly intermittent at low CR energies (GeV - TeV energies which we are interested in), the statistical description of the nonthermal particle distributions is different from that of the statistically-homogeneous systems \citep[][]{1990ZhETF..98.1255B}.  They decomposed the distribution function in two parts $F({\mbox{\boldmath $r$}},p,t) = N({\mbox{\boldmath $r$}},p,t) + \delta N({\mbox{\boldmath $r$}},p,t)$, where as usually  $<F({\mbox{\boldmath $r$}},p,t)> = N({\mbox{\boldmath $r$}},p,t)$ is the distribution function averaged over the ensemble of the large scale fluctuations.  Then at the high energy end $\eta(p) < $1  provides  $N({\mbox{\boldmath $r$}},p,t) >>  |\delta N({\mbox{\boldmath $r$}},p,t)|$ and the standard quasi-linear procedure of averaging is appropriate. For the low energy CRs where  $\eta(p) > $1 and the perturbation $|\delta N({\mbox{\boldmath $r$}},p,t)|$ is not small anymore at the shock fronts the perturbation theory fails, but one can use another approximation. In this case one can use a solution of the diffusion-advection equation which connects the particle distribution function at a given shock surface $ F_{\rm i}({\mbox{\boldmath r}},p,t)$ with the distribution function in the far upstream of the shock which can be approximated by the average distribution function  $N({\mbox{\boldmath $r$}},p,t)$. 

\begin{equation}
      F_{\rm i}({\mbox{\boldmath r}},p,t)  =  \theta(z_{\rm i})\Phi_{\rm i} +   \theta(-z_{\rm i}) \large[\Phi_{\rm i} \exp{(\Delta u_{\rm i}z_{\rm i}/\chi_{\rm i} z_{\rm i})} +  N(z_{\rm i},p)\large]
 \end{equation}
 where  $\theta(z_{\rm i})$ is the Heaviside step function of a coordinate $z_{\rm i}$ along the normal to the plane of the where ${\rm i}$-th shock surface with the flow velocity jump $\Delta u_{\rm i}$ and the particle diffusion coefficient in the shock upstream $\chi_{\rm i}$. The function  $\Phi_{\rm i}$ depends on the particle power law distribution of index $\gamma_{\rm i}$ which is formed by the diffusion shock acceleration:    
\begin{equation}
      \Phi_{\rm i}({\mbox{\boldmath r}},p,t)  =  (2 +\gamma_{\rm i}) p^{-(2 +\gamma_{\rm i})} \: \int_{0}^{p} {\rm d}p' \: {p'}^{(\gamma_{\rm i} +1)}N(p'). 
 \end{equation}

In the simplest case  $\gamma_{\rm i} =(\sigma_{\rm i} +2)/(\sigma_{\rm i} - 1)$ is
determined by the shock compression ratio $\sigma_{\rm i}$, while some other effects like the Alfvenic drift \citep[see e.g.][]{2018ApJ...856...33K} can 
be important and will .  

In this approximation \citet[][]{1990ZhETF..98.1255B} obtained the following system of equations for the distribution function  $N({\mbox{\boldmath $r$}},p,t)$  averaged over the ensemble of large scale plasma motions with frozen-in magnetic field fluctuations of a broad scale range and shocks: 

\begin{equation}
      \frac{\partial N}{\partial t} -
       \frac{\partial}{\partial r_{\alpha}} \: \chi_{\alpha \beta} \:
       \frac{\partial N}{\partial r_{\beta}}  =
       G  \hat{L} N +
      \frac{1}{p^2} \: \frac{\partial}{\partial p} (\: p^4 D  - \:p^2 \dot{p_L}\:)
      \frac{\partial N}{\partial p} + A {\hat{L}}^2 N +
      2B \hat{L} \hat{P} N + Q_{j}(p),
\label{KinEq}
\end{equation}

The operators  $\hat{P}$ and $\hat{L}$ are given by the equations

\begin{equation}
 \hat{P}= \frac{p }{3} \: \frac{\partial}{\partial p}\;;~~~~~       
\hat{L}= \frac{1}{3p^2} \: \frac{\partial}{\partial p} \:
      p^{3-\gamma} \: \int_{0}^{p} {\rm d}p' \: {p'}^\gamma
      \frac{\partial}{\partial p'}\, .
\end{equation}

The source term $ Q_{j}({\mbox{\boldmath $r$}},p,t)$ is determined by the properly averaged injection
rate of the nuclei of a type $j$, while  $\dot{p_L}(p)$ is the rate of 
particle energy losses. 
The kinetic coefficients $A$, $B$, $D$, $\tau_{sh}$, and
$\chi_{\alpha \beta}$
are expressed in terms of the spectral functions
which describe correlations between large scale turbulent motions and shocks. 

We presented here a simplified version of the kinetic equations assuming that all of the 
$\gamma_{\rm i} =  \gamma$ which is the case either for the ensemble of strong shocks 
(where the indexes are close to 2), or for the shocks of identical strengths. The general 
case of shock ensemble is described by equations presented in \citet[][]{1990ZhETF..98.1255B,1993PhyU...36.1020B} .  

The kinetic coefficients satisfy the following renormalization equations (assuming a quai-isotropic turbulent field):

\begin{equation}
  \chi = \kappa + {1 \over 3} \int { d^3 {\bf k} \, d\omega \over (2\pi)^4 }
  \left[ {2T+S \over i\omega+k^2\chi }
        -{2k^2\chi S \over \left( i\omega + k^2\chi\right)^2 } \right] \, ,
\label{Chi}
\end{equation}

\begin{equation}
  D={\chi \over 9} \int { d^3 {\bf k} \,  d\omega \over (2\pi)^4 }\;
   {k^4 S(k,\omega,t) \over \omega^2 + k^4 \chi^2 } \, ,
\label{D}
\end{equation}

\begin{equation}
 A = \chi \int { d^3 {\bf k} \, d\omega \over (2\pi)^4 } \;
   {k^4 \tilde{\phi}(k,\omega,t) \over \omega^2 + k^4 \chi^2 }\, ,
\end{equation}

\begin{equation}
  B = \chi \int { d^3 {\bf k}\, d\omega \over (2\pi)^4 }\;
  {k^4 \tilde{\mu}(k,\omega,t) \over \omega^2 + k^4 \chi^2 }\, .
\label{eq:B}
\end{equation}

Here  $T(k,\omega,t)$ and $S(k,\omega,t)$  are
the transverse and the longitudinal parts
of the Fourier components of the turbulent velocity correlation tensors. The time dependence 
is determining the slow evolution of the spectra on the time intervals longer $t > \omega^{-1}$.  
$G = ( 1/\tau_{sh}+ B)$ are determined by the shock statistics   \citep[see e.g.][]{1987Ap&SS.138..341B,ry03}. 
Correlations of the velocity jumps at shock fronts 
are described by $\tilde{\phi}(k,\omega,t)$, while $\tilde{\mu}(k,\omega,t)$
represents the correlations of shocks and long wavelength rarefaction waves. 
The introduction of these spectral functions is required due to the intermittent
character of the system with multiple shocks. The local  diffusion coefficient $\kappa(p) = v\Lambda(p)/3$ is due to the CR scattering by short scale magnetic fluctuations. At low energy regime under consideration where $\eta(p)  > $1 the local diffusion coefficient   $\kappa(p)$  is smaller than 
$\chi$, and the global  transport of the low energy CRs is governed by the turbulent diffision. 

As it is seen from Eq.\ref{D} in the case of Fermi acceleration of CRs by the long wavelength MHD turbulence 
(where the energy containing scale size of the turbulence $L$ is larger than the CR mean free path
$\Lambda(p)$) the compressible longitudinal part of the spectrum dominates the acceleration \citet[see for review][]{1993PhyU...36.1020B}. Therefore the Fermi type acceleration of CRs by the longitudinal large scale MHD motions result in a damping of the turbulent motions if the CR population is numerous enough.  On the other hand the turbulent diffusion of CRs in the large scale MHD turbulence as it follows from Eq.~\ref{Chi} is provided by both the incompressible vortex type motions 
(described by the transverse part of the spectrum  $T(k,\omega,t)$ ) and the compressible turbulence (longitudinal part of the spectrum $S(k,\omega,t)$). 

\subsection{Spectra of cosmic rays accelerated by ensemble of shocks}
\label{NLensemble}
The kinetic energy of the baryonic component of a large cluster of galaxies can be as large as $\sim 10^{63}$ ergs and the kinetic power release within the intercluster meduim may reach $10^{47}$ erg s$^{-1}$
during  major merger events. 
Vortex electric fields generated by the large scale motions of highly
conductive plasma with shocks result in a non-thermal distribution of
the charged particles. This may transfer some fraction of the kinetic power of the baryonic matter into the non-thermal components. The superthermal particles both freshly injected at shocks and the long lived pool  produced by the previous substructure mergers, AGN activity and the other processes  in clusters of galaxies can be reaccelerated by the large scale MHD motions of magnetized plasma with shocks.

We can apply the kinetic equation Eq.~(\ref{KinEq}) described above to model the efficiency of the conversion of the kinetic power of the compressible supersonic turbulence to CR production. The  model considers a simplified statistically homogeneous system, which only consists of the large-scale turbulent plasma motions and CRs, and the total energy of the system is conserved. In the model,  the spectra of turbulence were parameterized in the following way: 
\begin{equation}
\psi(k,\omega,t)=\psi(k,t){\Gamma_k \over 2}\left[
{{1 \over (\omega - \omega_0)^2 + (\Gamma^2_k /4)} +
{1 \over (\omega + \omega_0)^2 + (\Gamma^2_k /4)}}\right]\ .
\end{equation}
Here we introduced $\psi(k,\omega,t) = (T(k,\omega,t), S(k,\omega,t))$, 
$\Gamma_k=\tau_c(k)^{-1}$,  $\omega_0(k) = v_{\rm ph}k$, 
$S(k,t)\propto S_0(t)e^{-{k^2 \over k^2_0}}$, where $k_0= 2\pi L^{-1}$

\begin{figure*}
	\centering
    \includegraphics[width=0.99\textwidth]{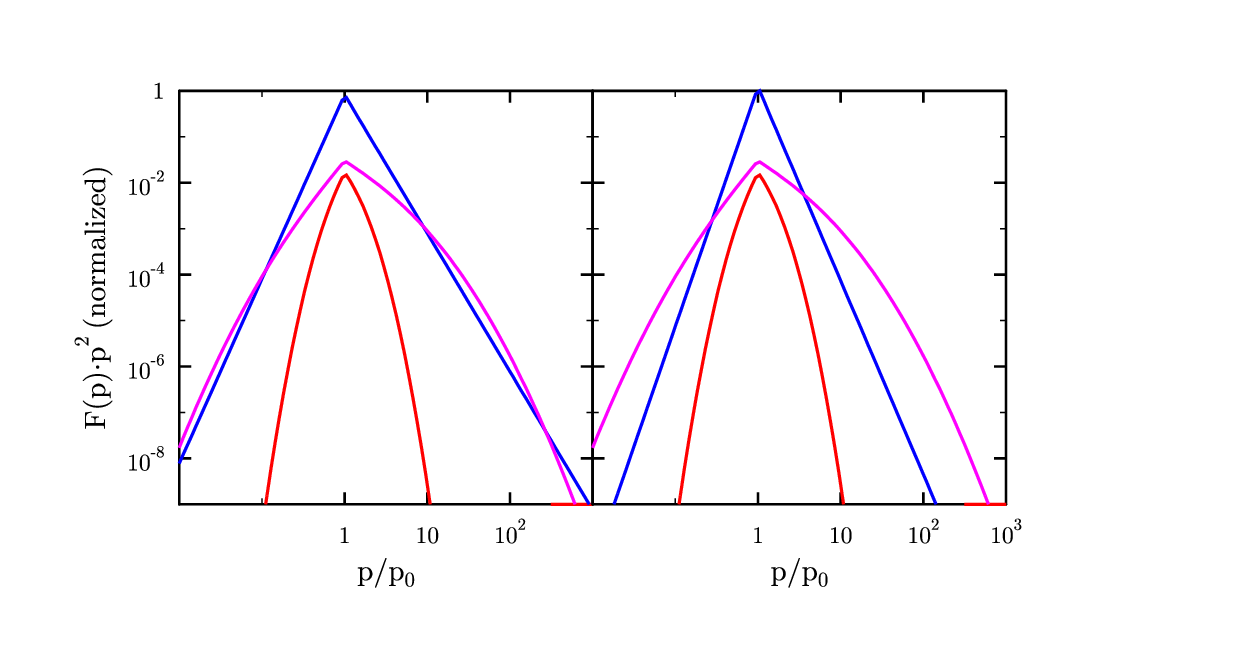}
    \caption{The temporal evolution of the nonthermal protons for a cluster of a size $\sim$ 2 Mpc with the large scale turbulence of $L \sim 100$ kpc and velocity magnitude 2$\times 10^3$ km/s. CRs were injected at $p=p_0$  with the energy density about 10$^{-4}$ (left panel) and 10$^{-3}$ (right panel) of that of the turbulence. The spectra are shown at the time moments -- 0.3  $\tau_{\rm a}$ (red curves), 1 $\tau_{\rm a}$ (magenta curves) and the asymptotic spectra after 3 $\tau_{\rm a}$ (blue curves).} \label{fig:Spectra}
\end{figure*}

In  Fig.~\ref{fig:Spectra} we illustrate the simulated temporal evolution of the spectra of CRs accelerated in a cluster of about 2 Mpc scale size with the amplitude of the bulk velocity fluctuations  $\sim$ 2,000 km/s (i.e. $\sqrt{S(0)k_0} \sim$ 2,000 km/s)  and the energy containing scale
$L =100$ kpc.  In this case we assumed  a reacceleration scenario where CRs had initially a narrow energy spectrum with the energy density of about 10$^{-4}$ (left panel) or 10$^{-3}$ (right panel) of that of the turbulence. The conversion efficiency could  exceed 10\%. The asymptotic CR energy spectra at the time moments after 3 $\tau_{\rm a}$, where $\tau_{\rm a} = 9L/u$  are approaching  the power-law with $N(p)p^2 \propto p^{-a}$ (blue curves). The asymptotic spectra are harder in case of the lower injection rate. At earlier time moments 0.3  $\tau_{\rm a}$ (red curves) and 1 $\tau_{\rm a}$ (magenta curves) one can see some spectral evolution from soft to hard  and then asymptotically to the power-law spectra. 
 The model is simplified, yet it can provide an illustration of the possible role of non-linear feedback on the spectral evolution. More exact models must account for the evolution of the short scale collisionless turbulence which provide the CR scattering ($\Lambda(p)$  was fixed here) and a consistent model of CR injection and  gas heating by shocks as it was discussed in \S\ref{PICshock}.

\subsection{Particle acceleration by multiple shocks in the heliosphere}

Cosmic ray acceleration by the ensemble of shocks and long wavelength MHD motions are likely to occur in other astrophysical objects. Direct measurements of both magnetic turbulence and non-thermal particles are available in the heliosphere. The analysis of the suprathermal spectra in the interplanetary medium observed in the quiet periods revealed the spectra $f(v) \propto v^{-5}$ \citep[][]{2006ApJ...640L..79F,2014JGRA..119.8733F,2017JPhCS.900a2006F}. The spectra are consistent with the model for the intercluster medium $N(p)p^2 \propto p^{-3}$ as in the left panel in  Fig.~\ref{fig:Spectra} (see also Fig.~2  in \citep[][]{2001SSRv...99..317B}). The origin of the spectra observed  in the interplanetary medium is associated with  particle acceleration by the compressional plasma wave trains consisting of a series of compression or rarefaction regions \citep[][]{2006ApJ...640L..79F,2010JGRA..11512102Z,2010ApJ...713..475J,2013ApJ...764...89A,2017JPhCS.900a2006F}
which can be described by Eqs.~(\ref{KinEq})- (\ref{eq:B}) and spectra given in Fig.~\ref{fig:Spectra}. It is important that in this case observations revealed the extended spectrum of magnetic fluctuations which provide effective scattering of non-thermal particles.

\subsection{Particle acceleration by multiple shocks in galactic superbubbles}

Energetic particles likely comprise a long-lived CR component which may contain a sizeable fraction of the kinetic energy released by the winds and supernova ejecta in galactic superbubbles \citep[][]{2001SSRv...99..317B,2014A&ARv..22...77B}. Strong supernova shocks accompanied by the multiple secondary weak shocks may be the responsible for 
the observed excess of gamma-ray emission detected by Fermi-LAT in the Cygnus OB2 association  
\citep[][]{AckermannSB2011} and in the star forming region
G25.0+0.0 \citep[][]{Katsuta2017}. The observed gamma-ray spectra  roughly correspond  to CR  distribution $N(p)p^2 \propto p^{-2}$. The hard spectra are expected to be the case in the systems with strong shocks like those produced by supernovae  \citep[][]{2001SSRv...99..317B,2014A&ARv..22...77B,Grenier2015,Lingenfelter17,2018SSRv..214...41B}.  
In the clusters of galaxies with CR acceleration by weak shocks and long-wavelength turbulence in the model discussed above the spectra are softer with $N(p)p^2 \propto p^{-3}$ if the broad band of magnetic fluctuations providing the effective CR scattering is available in the intercluster medium. 
However, if the scattering of CRs in GeV-TeV regime is inefficient and $\Lambda(p)$ is long enough 
to provide $\eta(p) <<1$, then the CR acceleration time is getting longer than their escape time and the spectra are steeper.  Observations of the non-thermal emission \citep[see e.g.][]{2008SSRv..134...93F,2012A&ARv..20...54F,bj14,2014A&A...567A..93P,2015A&A...578A..32Z} as well as the future high resolution X-ray spectroscopy \citep[][]{2009A&A...503..373K} can constrain the particle acceleration scenario and the contribution of the non-thermal components to the energy budget of the clusters.

Apart from the diffusive CRs propagation in clusters of galaxies and the diffusive shock acceleration which is considered as a rather realistic mechanism of the efficient acceleration of non-thermal particles in clusters of galaxies the superdiffusive propagation of CRs may also be essential under some conditions in both accretion and merger shocks \citep[see e.g.][]{Bykov_SuperD2017,2018MNRAS.478.4922Z} and the magnetic reconnection process was discussed in this context by  \citet[][]{2011MmSAI..82..636L}.

\section{Cosmological simulations of cosmic ray acceleration and their problems}
\label{CS}

Because of the expected complexity of the shock network in the dynamical environment of galaxy clusters, following the spectral and spatial evolution of accelerated CRs in the intracluster medium is a nontrivial task. The acceleration of cosmic ray electrons and of cosmic ray protons in cosmology generally poses very different challenges to numerical simulations, and for this reason the two processes are usually followed by means of two complementary approaches. 

\subsection{CR electrons in cosmological simulations}
Unless they get (re)accelerated multiple times, cosmic ray electrons are expected to be short lived in typical ICM condition  ($\leq 10^8 ~\rm yr$  for radio emitting  $\gamma \sim 10^4$ electrons, e.g. Van Weeren et al. this topical collection). Once accelerated, they  cannot thus be  advected nor diffuse more than $\sim 100 ~\rm kpc$ away from their injection site,  before they loose most of their energy via synchrotron and Inverse Compton radiation. Given the energy budget which can be derived by radio observations, their dynamical impact on the ICM is also believed to be entirely negligible.

These circumstances justify the simple post processing treatment of CR electrons in most cases: typically once that a cosmological simulation is run, shock waves are detected using different shock finder methods, and the budget of CR electrons accelerated in the shock downstream is computed based on the information of a single timestep, assuming a quasi-stationary balance of DSA acceleration and radiative losses \citep[e.g.][]{hb07}. 

\begin{figure*}
	\centering
    \includegraphics[width=0.32\textwidth,height=0.32\textwidth]{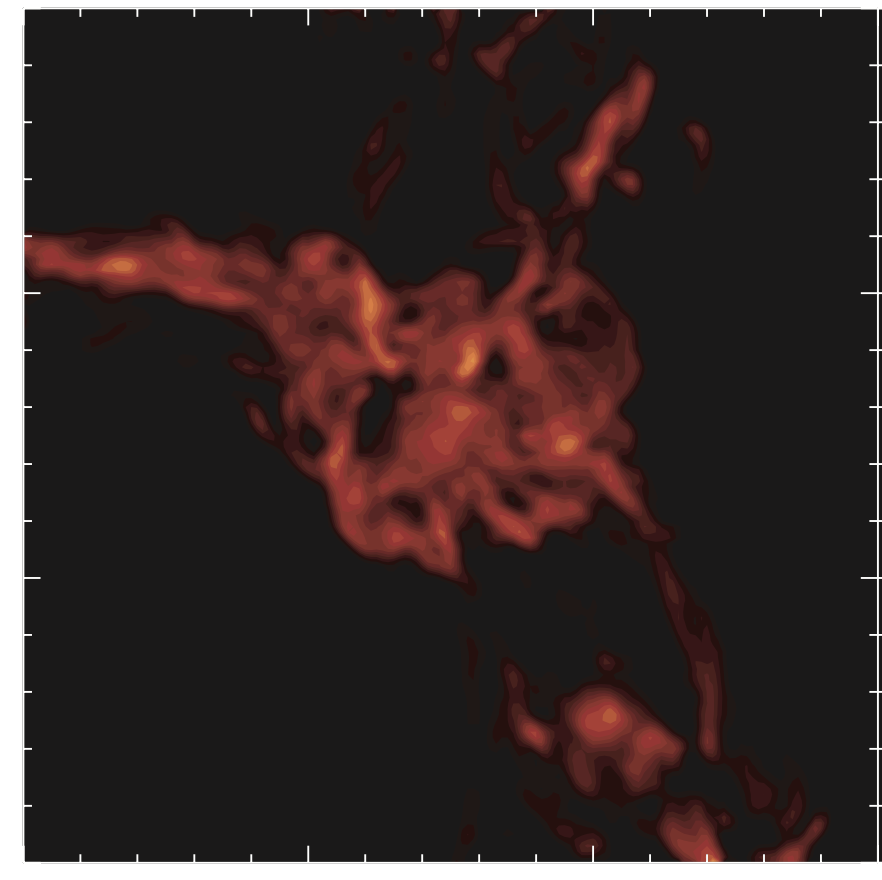}
    \includegraphics[width=0.32\textwidth,height=0.32\textwidth]{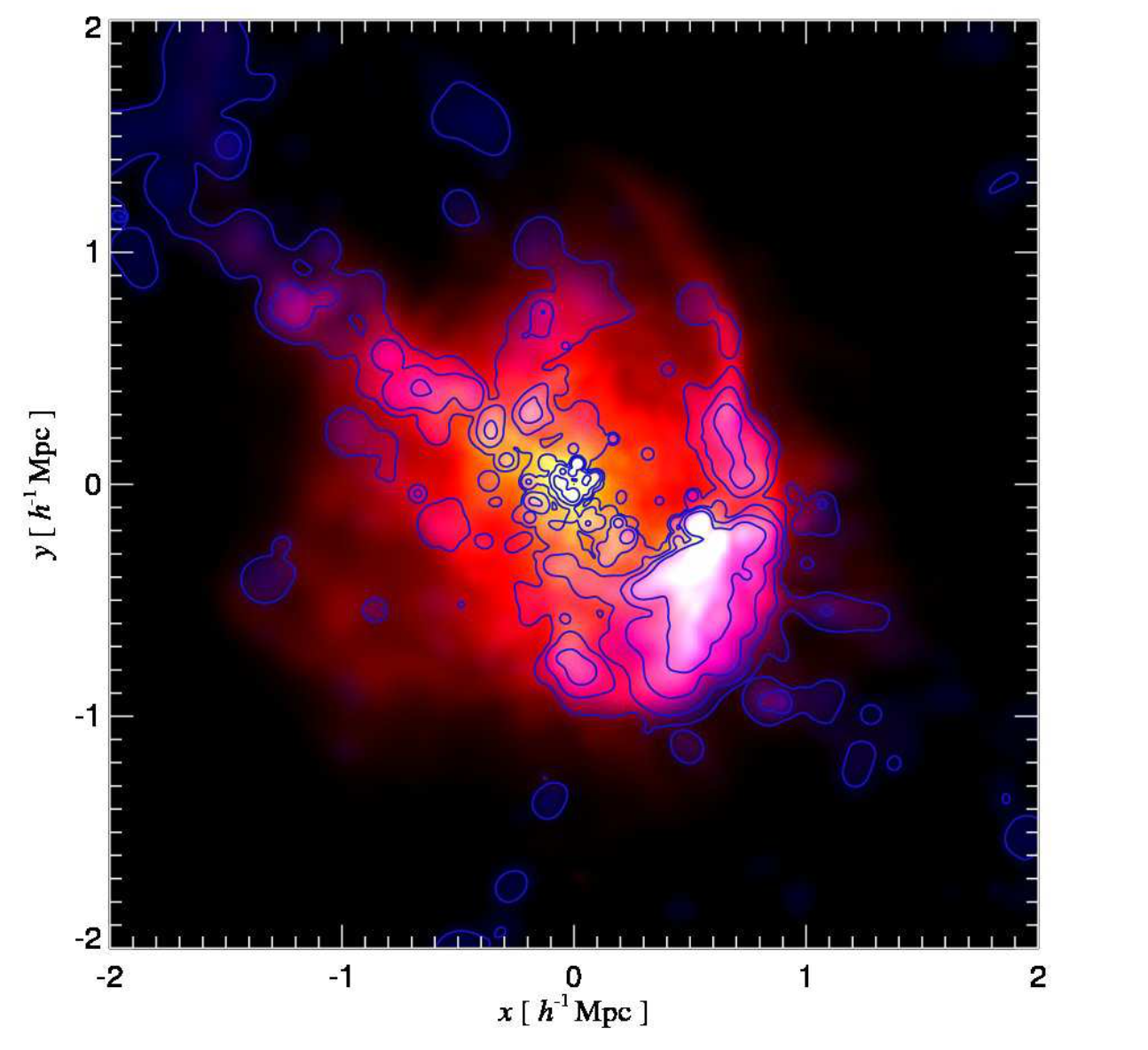}
    \includegraphics[width=0.32\textwidth,height=0.32\textwidth]{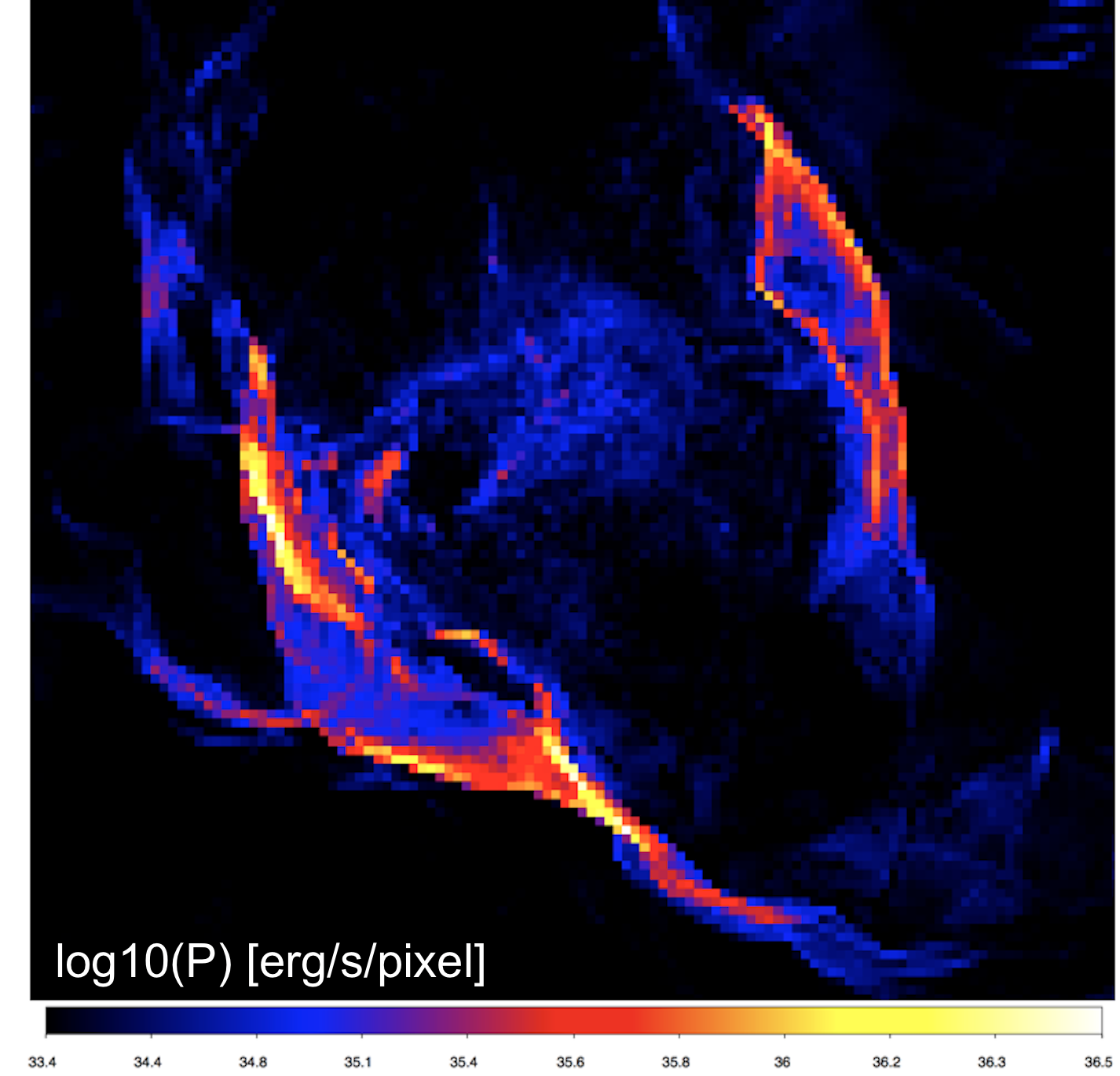}
    \includegraphics[width=0.49\textwidth,height=0.45\textwidth]{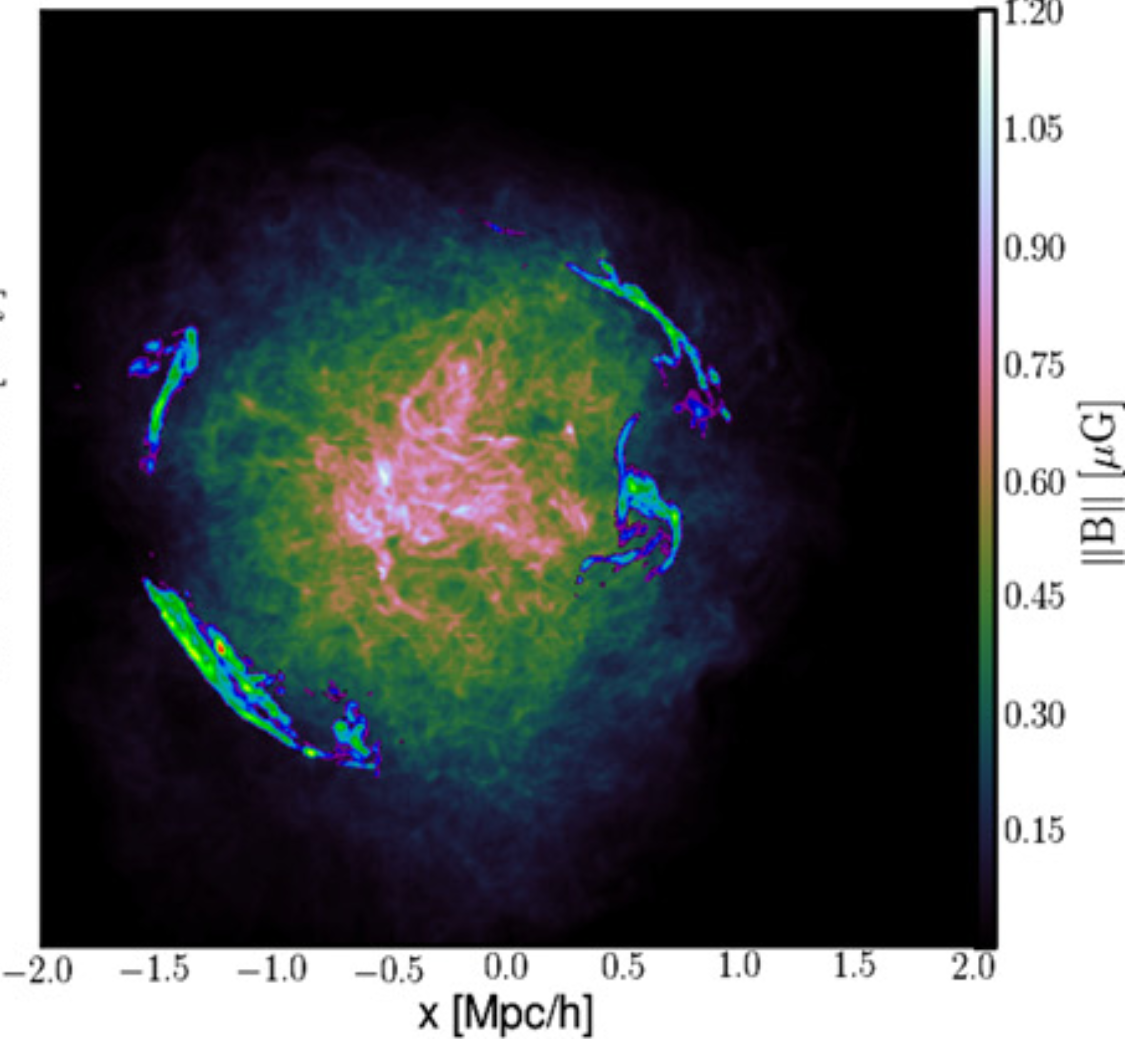}
    \includegraphics[width=0.49\textwidth,height=0.45\textwidth]{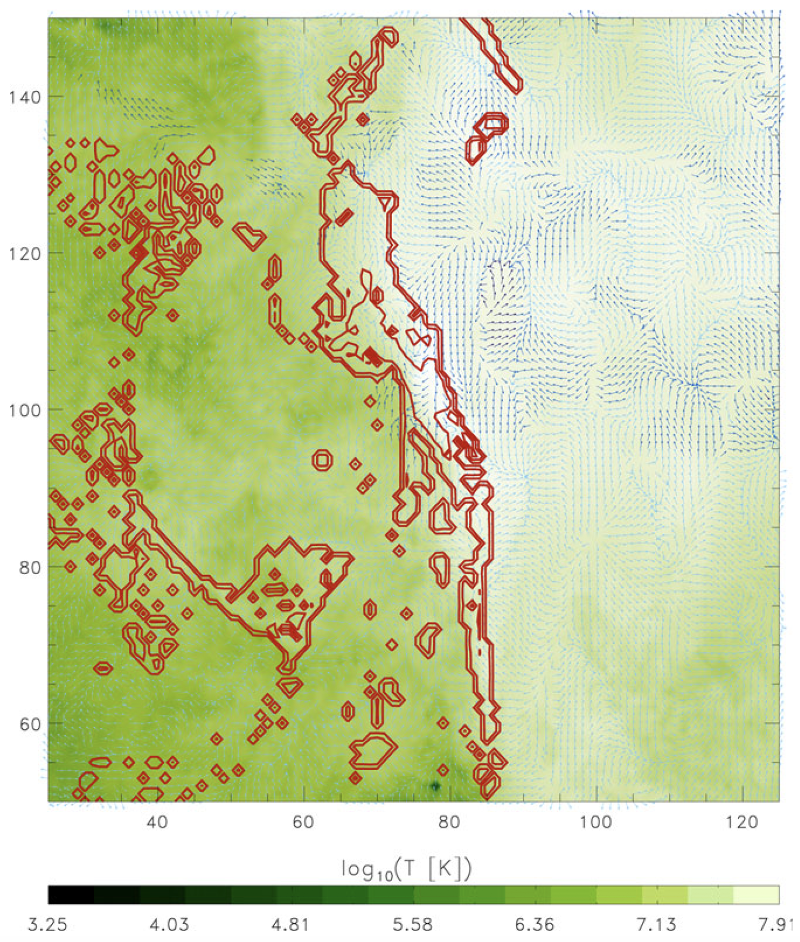}
    
    \caption{Examples of simulated non-thermal emission from shock accelerated electrons in cosmological simulations. Top left panel: Hard-X emission ($\geq 100 ~\rm keV$) from the Inverse Compton scattering of electrons injected by cosmic shocks around a galaxy clusters, simulated with a fixed mesh simulation by \citet{2003MNRAS.342.1009M}.  Top central panel: primary radio emission at $150 ~\rm MHz $ (the contours show the emission detectable above $7 \cdot  10^{-4} \rm  ~mJy/arcmin^2$) in an SPH simulation by \citet{2009MNRAS.393.1073B}, with a post-processing magnetic field which scales with thermal gas energy.  Top right panel: radio emission from shock-accelerated emission in an Eulerian AMR simulation of a galaxy clusters (the image is $5~\rm  Mpc/h$ across) by \citet{va12relics}, for an assumed flat magnetic field of $5 \mu G$ everywhere.  Bottom left panel: simulated radio emission at $1.4 ~\rm GHz$ from shock-accelerated electrons (bright colors) in the Eulerian AMR simulation by \citet{sk13}, featuring a run-time simulated magnetic field injected by past AGN activity (color map). Bottom right panel: close up view of a radio relic in a galaxy clusters simulated using Eulerian AMR with run-time magnetic fields (assumed of primordial origin) by \citet{wi17}, and imposing that only quasi-perpendicolar shocks efficiently accelerate electrons. The zoom covers $\sim 3.2 \times 3.2 \rm ~Mpc^2$, the color map shows the projected gas temperature, the red contours the detectable radio emission at $1.4 ~\rm GHz$, and the vectors show the local direction of magnetic fields.} \label{fig:jub}
    \label{fig:relics_sim}
\end{figure*}

Very first works that studied the acceleration and energy evolution of electrons in the test particle regime, employed Eulerian schemes to model shocks and compute non-thermal emission from electrons \citep[e.g.][]{2001ApJ...562..233M,2003MNRAS.342.1009M}. Later on high resolution simulations employing smoothed-particle hydrodynamics \citep[][]{2004ApJ...617..281K,ho08,2008MNRAS.385.1211P, 2009MNRAS.393.1073B,2012MNRAS.423.2325A,2012MNRAS.420.2006N} or adaptive mesh refinement \citep[][]{va10kp,sk11,va12relics,2018MNRAS.476.4629P} modelled the radio emission from shock-accelerated electrons as well, assuming DSA acceleration and ad-hoc prescriptions on ICM magnetic fields in a post-processing step. A few examples of non-thermal emission from shock-accelerated electrons in cosmological simulations are given in Fig.\ref{fig:relics_sim}. 

The most direct testbed for cosmological simulations of electron acceleration is represented by the fast growing class of radio relics sources (see Van Weeren et al., this topical collection). 

\begin{figure*}
	\centering
    \includegraphics[width=0.49\textwidth]{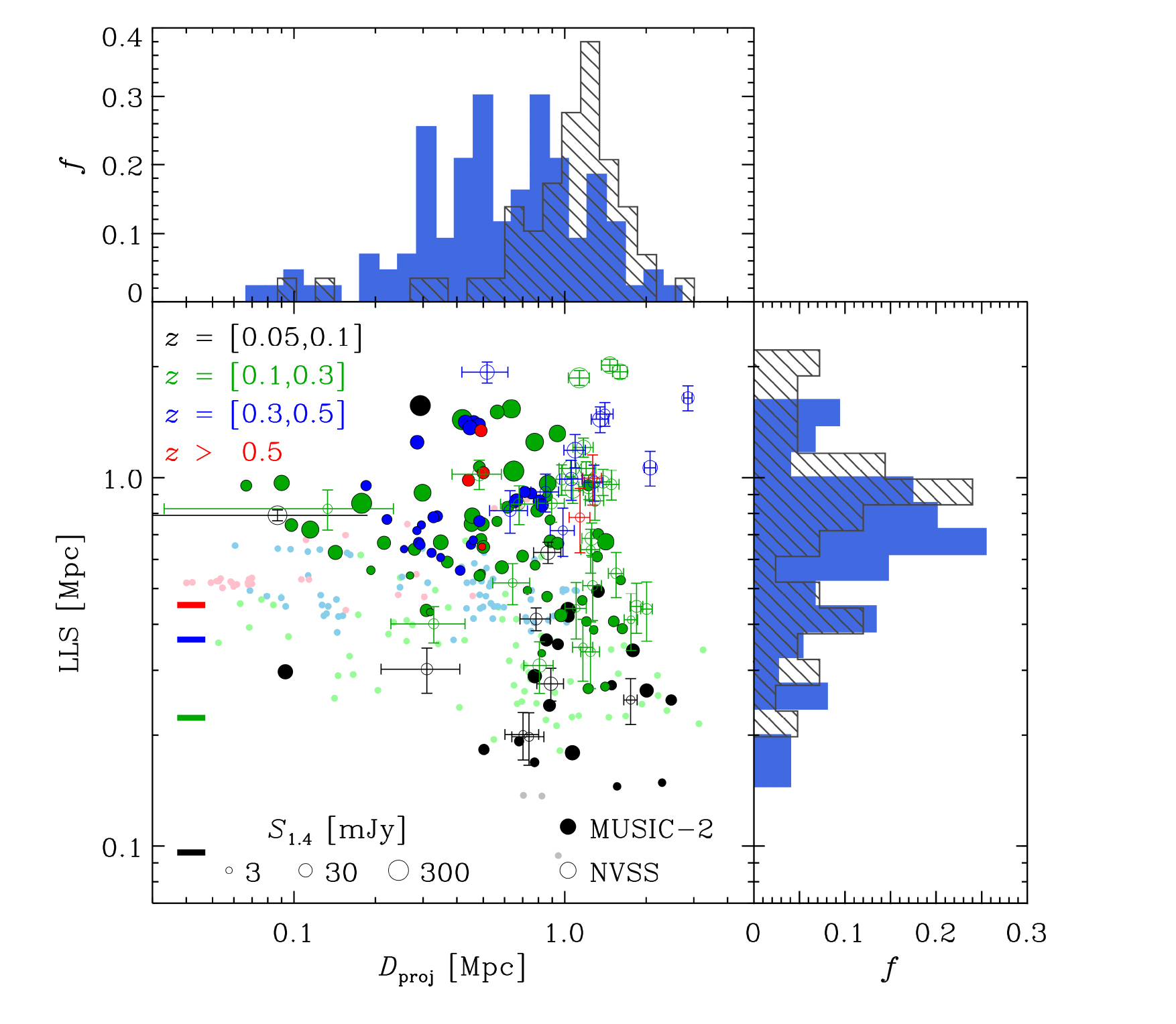}
    \includegraphics[width=0.49\textwidth]{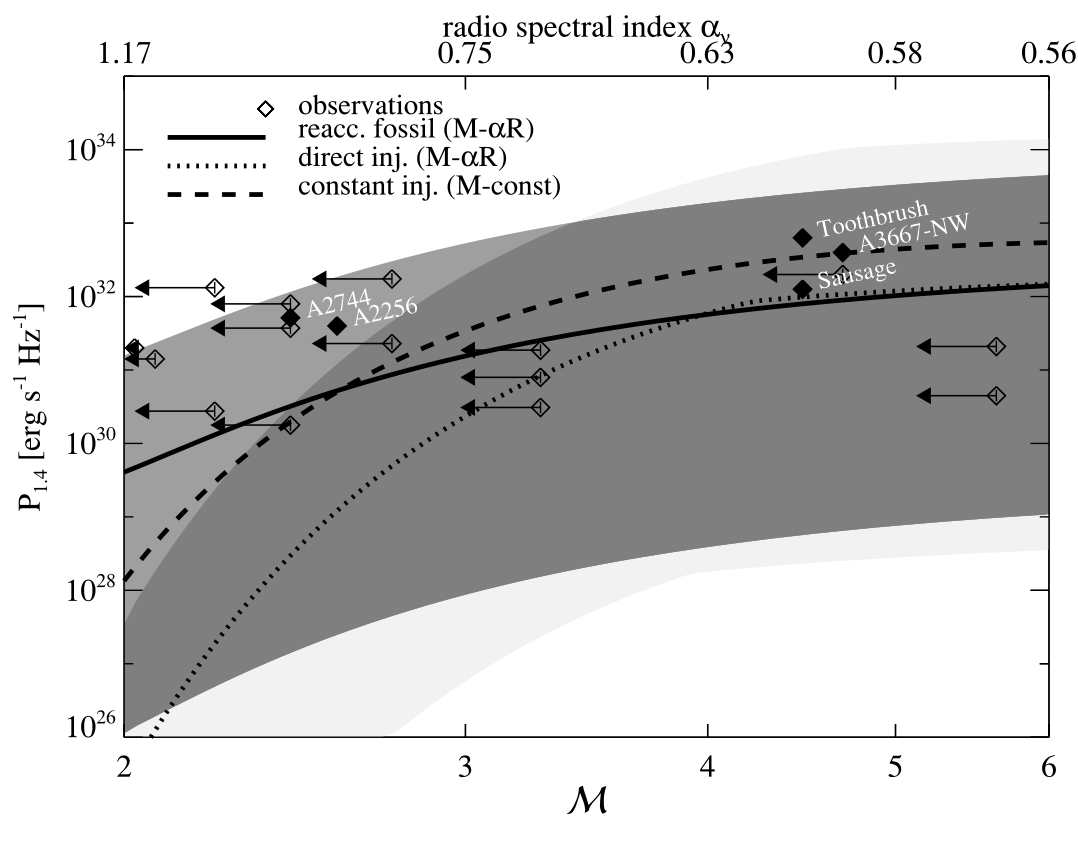}
    \caption{Left panel: comparison between the projected distance and the largest linear scale (LLS) of observed (filled circles) and simulated (open circles with different colors for different redshift) radio relics, from  \citet[][]{2017MNRAS.470..240N}.  Right panel: relation between the relic radio power  and the shock  Mach number for observed relics (diamonds,  with uncertainties given by the length of horizontal arrows) and reference acceleration models (with thickness of the areas given by the typical range of uncertainties  in the shock thickness, magnetic field, and downstream temperature and density) for three models: a model with fixed acceleration efficiency ("constant inj."), a model based on the DSA acceleration of fresh electrons only ("direct inj.") and a model including shock re-accelerated electrons ("reacc.fossil"). Adapted from \citep{2013MNRAS.435.1061P}.}
  \label{fig:nuza}
\end{figure*}

In this respect, cosmological simulations can in general produce simulated radio relics that are similar in shape, total power and spectral index to those observed in radio relics, suggesting that at first order the typical kinetic power from merger shocks in simulations is realistically the main source to power radio emitting electrons \citep[][]{ho08,2009MNRAS.393.1073B,sk11,2012MNRAS.426...40B,va12relics,2015ApJ...812...49H}. Also the
relation between the observed relic power and the host cluster mass or luminosity is reasonably close to observations \citep[][]{sk11,2012MNRAS.420.2006N,fdg14,2017MNRAS.470..240N}.
However, once large samples of observed and simulated radio relics are compared using the same analysis pipeline, important differences emerge in the distribution of their morphological parameters  \citep[][]{2017MNRAS.470..240N}: in particular simulations tend to find more "small" and roundish  relics than observations, and simulated relics are  found on average to be less extended (by $\sim 15 \%$ )and less distant (by $\sim 30\%$) from the cluster centre than their real counterparts in the NVSS survey (see left panel of Fig.\ref{fig:nuza}).  This may indicate that non-gravitational processes, like radiative cooling and energy feedback from galaxy evolution processes (not included in most simulations investigating radio relic emission), may affect the large-scale properties of radio emitting shocks as they sweep through the ICM. 

The "microphysical" uncertainties in simulations are still very large, and mostly reflect the complexity of plasma processes reviewed in the previous Sections. 
Simulations are bound to guess the acceleration efficiency by shocks in the $\mathcal{M} \leq 2-5$ regime, based on DSA or its extrapolation, and in the literature typical values from a few percent to  $\sim 10^{-5}$ of the shock kinetic energy are quoted to produce realistic radio power from merger shocks. 
It shall be noticed that acceleration efficiencies $\geq 10^{-3}$ poses problem to DSA as they imply an energy ratio of electron to proton at odds with basics expectations \citep[e.g.][ see also next Section]{va14relics,bj14}. 

Second, some degree of electron re-acceleration by shocks seems to be often necessary to explain radio emission from shocks with $\mathcal{M} \leq 2$ (see Van Weeren et al. this topical collection), which is larger than that from simple DSA (See right panel of Fig.\ref{fig:nuza}. 
This calls  for the modelling of older populations of electrons which may get re-accelerated by shocks coincident with relics \citep[e.g.][]{2013MNRAS.435.1061P}, and whose radio contribution may easily dominate over the one by "freshly injected" electrons for $\mathcal{M}\leq 4$ shocks \citep[][]{2012ApJ...756...97K}. Even more numerically challenging is the inclusion of turbulent re-acceleration of CR electrons in galaxy clusters \citep[][]{2013MNRAS.429.3564D, 2014MNRAS.443.3564D} which has so far never been obtained for fully cosmological simulations.  

Finally, only a few cosmological simulations to date could model the acceleration of relativistic electrons by shocks also including magnetic fields at run time, via MHD computations. 
\citet[][]{sk13} presented the first MHD grid simulation of radio relic emission in a galaxy cluster simulated at high resolution with Adaptive Mesh Refinement in the ENZO code, reporting a high degree of polarization, tentatively in agreement with observations available at the time, as well as the evidence of  a complex distribution of Mach numbers (and injection spectral indexes) at merger shocks (lower left panel in Fig.\ref{fig:relics_sim}). 
\citet{wi17} also recently used ENZO-MHD simulations and AMR to assess the distribution of shock obliquities in merger shocks (lower right panel in Fig.\ref{fig:relics_sim}), and studied the effect of a revised acceleration efficiency which get maximum for electrons at quasi-perpendicular shocks, as suggested by PIC simulations \citep{guo14a}, and in relation to the lack of hadronic emission from accelerated CR protons in galaxy clusters (see next Section).

\subsection{CR protons in cosmological simulations}
Different than in the case of CR electrons,  the simulation of CR protons in cosmology is bound to follow the acceleration and advection of particles in a more self-consistent way and at run-time, due to the very large ($\sim 1-10$ Gyr) lifetime of injected CRs.  
Several methods 
(mostly based on variations of "two-fluids models) have been developed over the last decade in order to couple the simulated evolution of large-scale structures to the accumulation of cosmic rays injected by cosmic shocks over time. 
While a few seminal works first explored the evolution of cosmic rays in cosmological grid simulations, neglecting the dynamical feedback of cosmic rays onto baryons \citep{mi01,2003ASPC..301..327R,2003MNRAS.342.1009M}, the first high-resolution simulations incorporating the dynamical impact of cosmic rays were based on  smoothed-particle hydrodynamics  \citep{pf07,en07,ju08,2008MNRAS.385.1211P,pi10}. \\

The above works have led to the first, quantitative predictions of the expected radial distribution of cosmic ray pressure relative to the gas pressure (see Fig.\ref{fig:jub}), as well as to the predictions on the level of $\gamma$-rays and radio emission from the hadronic mechanism. Cosmic rays accelerated by structure formation shocks were suggested to be volume filling, capable to significantly affect the gas dynamics in the cluster volume, and potentially responsible for the diffuse radio emission in the form of radio halos (\citealt{2008MNRAS.385.1211P}, see also  Van Weeren et al, this book).
Injected CR protons were also reported to  significantly reduce the star formation efficiency in small galaxies, as well as to affect the total mass-to-light ratio of small halos at the faint-end of the luminosity distribution \citep[][]{ju08}. Cosmic rays injected in galaxy clusters should  indeed lower the  effective adiabatic index of the gas $\Gamma_{\rm eff} \leq \Gamma=5/3$, and  within cool core regions the pressure from cosmic rays was found to approach equipartition with the thermal pressure, thereby enhancing the gas pressure and density and  affecting the  SZ signal \citep[][]{pf07}. In general, the most resolved SPH simulations reported on average $\sim 5\%$ CRs to thermal gas energy ratio within $R_{\rm vir}$ of clusters simulated with a non-radiative setup, and $\sim 10-30 \%$ with the inclusion of radiative gas cooling, with little dependence on the cluster dynamical state \citep[][]{ju08}. At the epoch in which such simulations were performed, their predictions resulted to be below the available {\sl  Fermi LAT}  upper limits, referred to $\sim 2.5$ years since the start of the mission (Fig.\ref{fig:fermi}).  \\

\begin{figure*}
	\centering
    \includegraphics[width=0.7\textwidth]{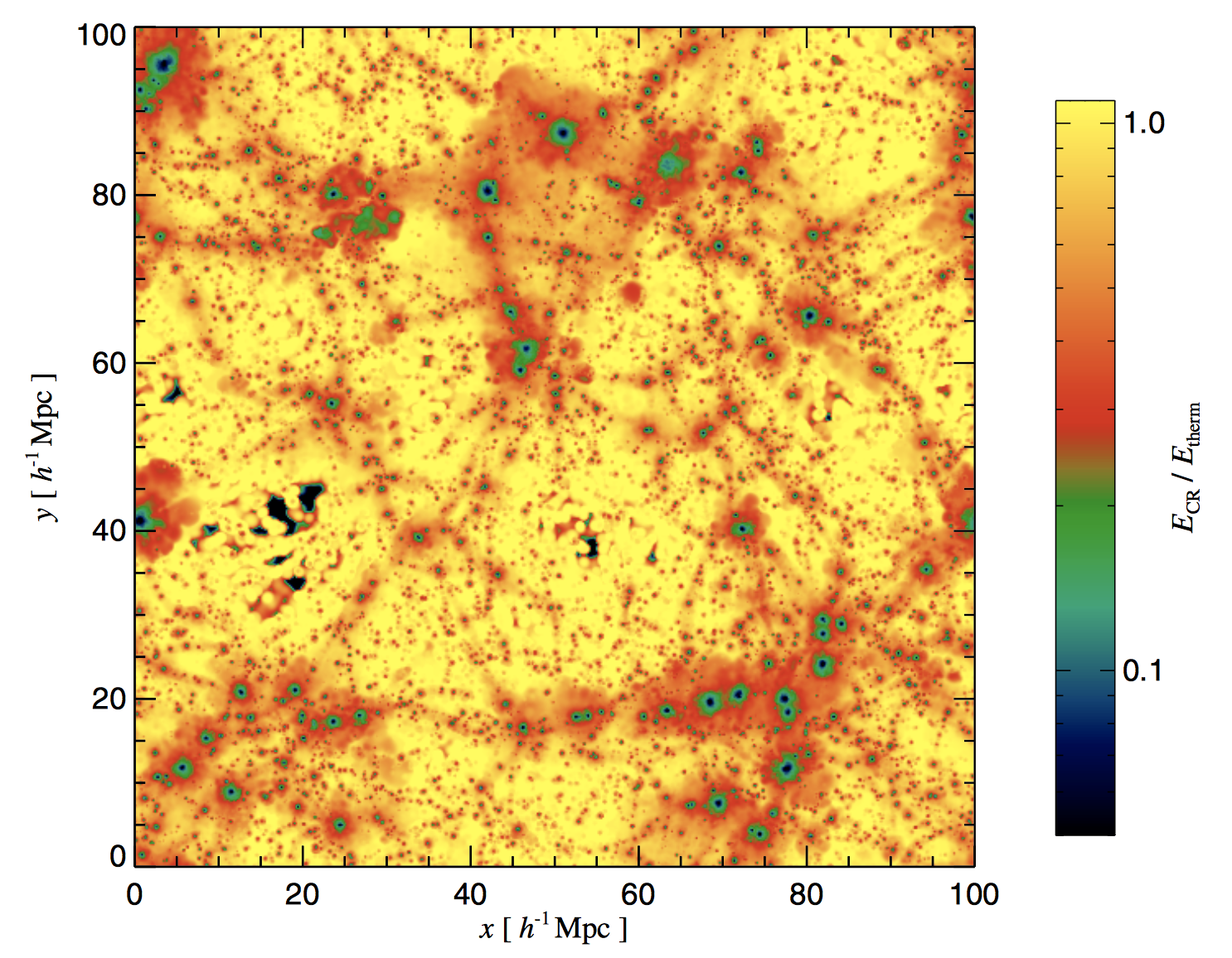}
    \caption{Projected maps of the pressure ratio between cosmic ray protons and thermal gas, in an SPH cosmological simulations by \citet{ju08}.} \label{fig:jub}
\end{figure*}

\begin{figure*}
	\centering
\includegraphics[width=0.95\textwidth]{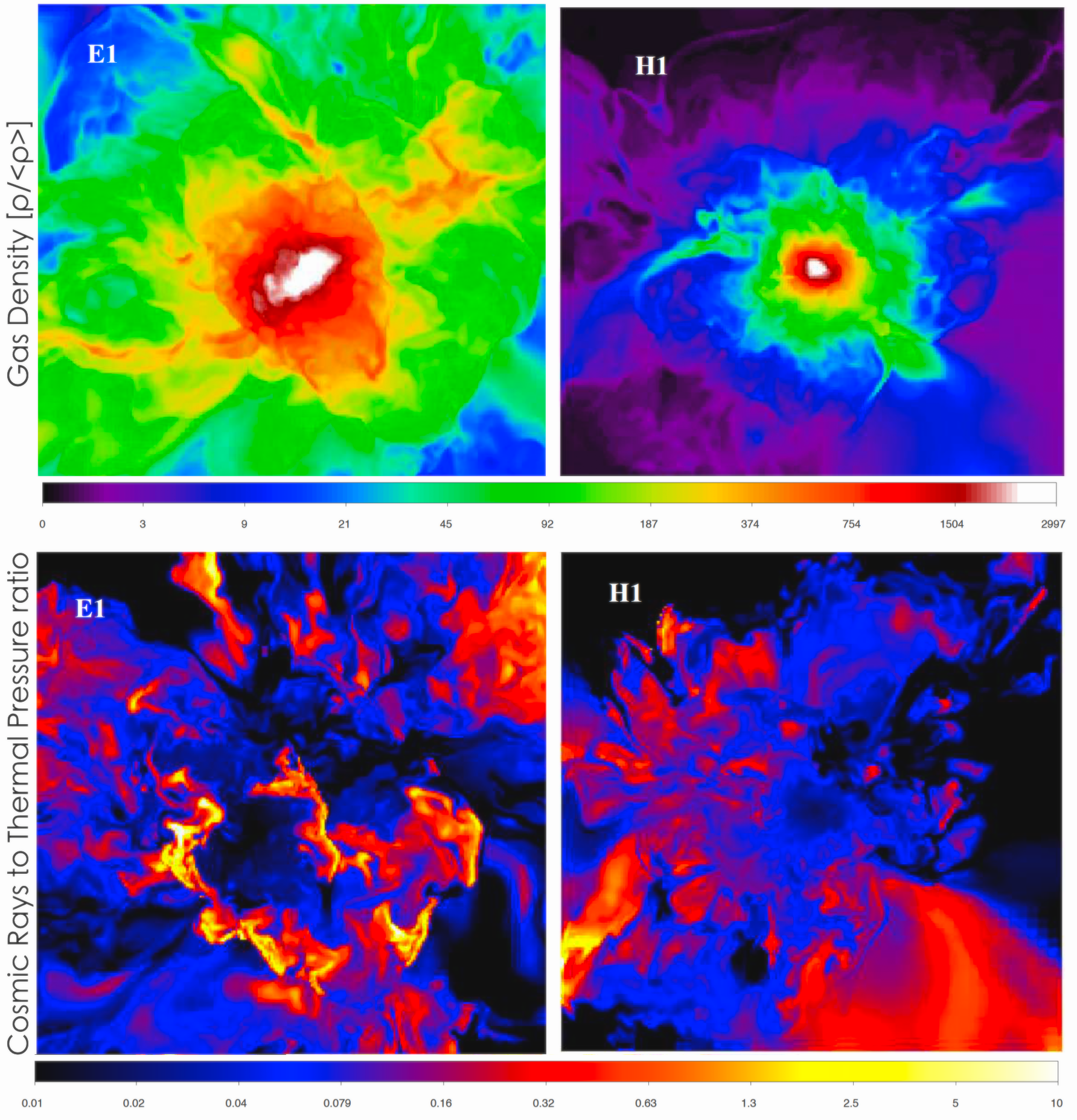}
    \caption{Gas density map (top) and pressure ratio between cosmic ray protons and thermal gas (bottom) for thin slices through the center of two galaxy clusters simulated with the {\enzo} grid code by \citet{scienzo}. } \label{fig:enzo}
    \end{figure*}

Later on, cosmological grid simulations (Fig.\ref{fig:enzo}) reached the necessary spatial detail to model the injection and advection of CRs with a two-fluid approach as well \citep[][]{scienzo,va14curie}, testing more modern versions of diffusive shock acceleration, and also including non-gravitational gas physics and shock re-acceleration physics \citep[][]{va13feedback,scienzo16}. 
A few key  differences were found respect to SPH results, ascribed to the different way in which the advection and entropy mixing of the two methods are numerically handled \citep[e.g.][]{ag07,va11comparison}, as well to the fact that these more recent simulations adopted revised version of the diffusive shock (re)acceleration efficiency by \citet{kj07} and \cite{kr13}.  As most of these works were performed after the report of the latest {\sl  Fermi LAT}  upper limits \citep[][]{fermi14}, they could directly address the issue of limiting the acceleration efficiency of CR protons from the comparison with observations, as also most of the early grid simulations predicted hadronic fluxes in excess to observations \citep[e.g.][]{va13feedback}.\\

\begin{figure*}
	\centering
     \includegraphics[width=0.9\textwidth]{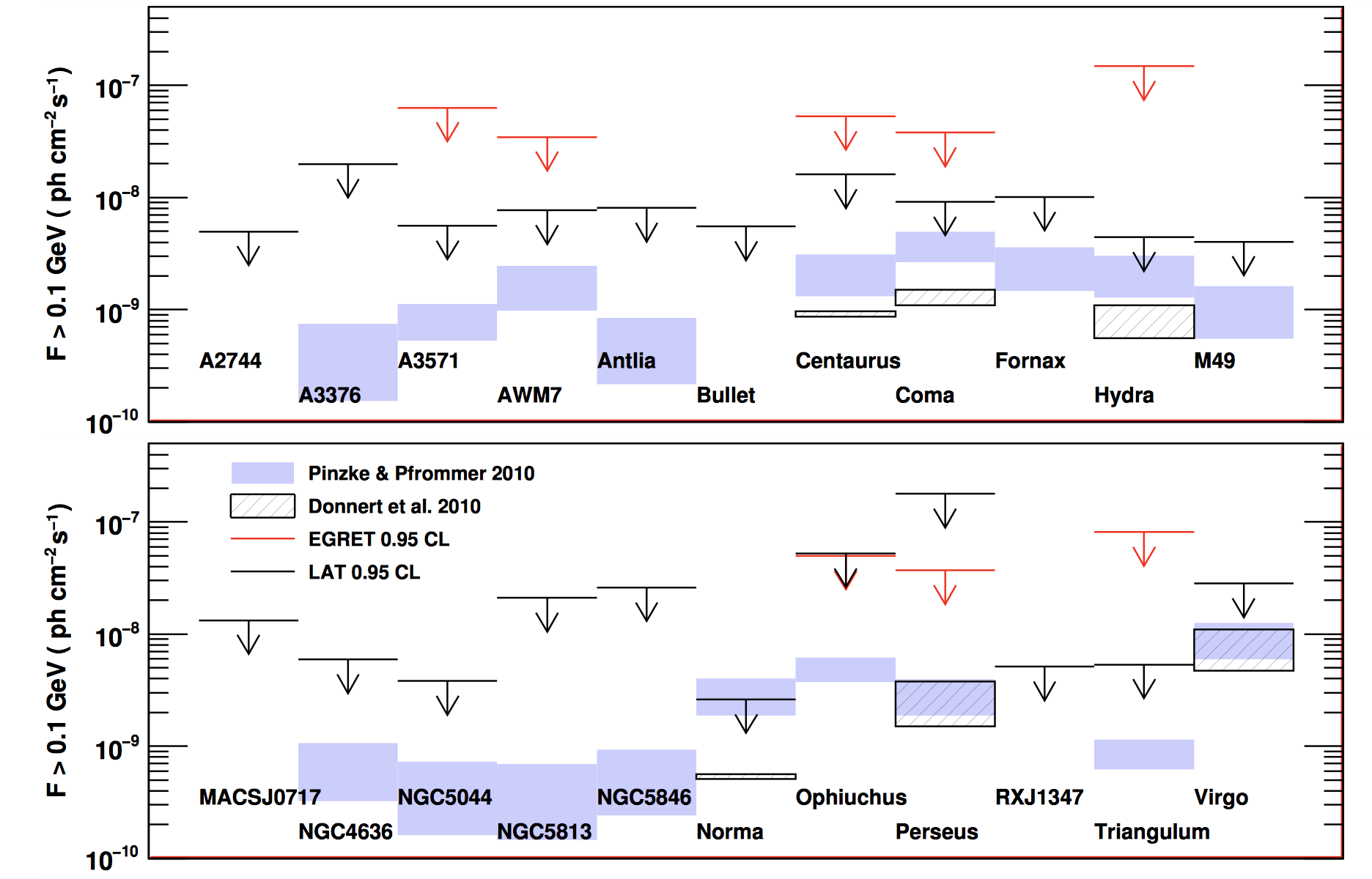}
    \caption{ Photon flux upper limits derived from Fermi-LAT by \citet{ack10} for the observations of several galaxy clusters (assuming unresolved gamma-ray emission) are compared to EGRET results 
   \citep[][]{re03}. The rectangles report the predictions based on SPH simulations using a two-fluid model  \citep[][]{pi10} and SPH simulations using a semi-analytical modeling of cosmic ray protons and radio emission \citep[][]{donn10}.  Taken from \citet{ack10}.} \label{fig:fermi}
\end{figure*}

In most tested models,  the predicted hadronic $\gamma$-ray emission is in excess to {\sl  Fermi LAT}  limits, for a $\sim 30-60\%$ of simulated clusters  \citep[][]{scienzo16}. The inclusion of cooling and feedback was found to significantly worsen the problem {\footnote{We notice that 
more recent works also included the effects of CR diffusion  \citep[][]{2016MNRAS.456..582S,2018MNRAS.475..570J}, and CR streaming \citep[][]{2017MNRAS.467..906W} in affecting the launching of star formation wind and their impact on the surrounding medium, even if the role of this in the total energy budget of CR in the intracluster medium on $\sim ~\rm Mpc$ scales is not significant. }}.  At variance with previous SPH works, Eulerian simulations reported a larger variance of the CR  to gas pressure within the virial volume of clusters with the same mass, following from an overall stronger dependence of clusters' dynamical history and the injection and re-acceleration of CRs.   
 If this is the case, the $\sim 1\% $ upper limits on the ratio between CR energy and thermal gas energy that can be derived from {\sl  Fermi LAT}  should mostly constrain galaxy clusters with a highly disturbed dynamical state, while in the rest of more relaxed systems the energy ratio within the virial volume may be a factor $\sim 10$ smaller, i.e. $\sim 0.1 \%$ \citep[][]{scienzo16}.   
As an example, the left panel of Fig.\ref{fig:ratio1} shows the predicted $\gamma$-ray emission level for clusters simulated in \citet{scienzo16} and contrasts it with the {\sl  Fermi LAT}  limits  \citep{fermi14}. 
In the baseline scenario the authors tested the 
(re-)acceleration efficiencies derived by \citet{kj07} or \citet{kr13}, which are based on the outcome of 1-dimensional diffusion-convection equation of cosmic shocks, and allows to study the effect of simplistic acceleration efficiency that simply scales with the shock Mach number. 
Among the tested models, only a model  in which the shock (re)acceleration efficiency of CR protons is bound to the low efficiency floor of $\eta=0.1\%$ (in the very simplistic assumption that the acceleration efficiency is independent on the shock Mach number) would produce $\gamma$-ray emission below {\sl  Fermi LAT}  limits. This corresponds to a very flat profile for the pressure ratio between accelerated CRs and thermal gas in most galaxy clusters at $z=0$, giving $\leq 0.5\%$ within $R_{\rm 200}$ of most objects. However, such a small acceleration efficiency for CR protons poses in several  cases problems to the "simple" explanation of radio relics with DSA (see below). 

\begin{figure*}
	\centering
 \includegraphics[width=0.7\textwidth]{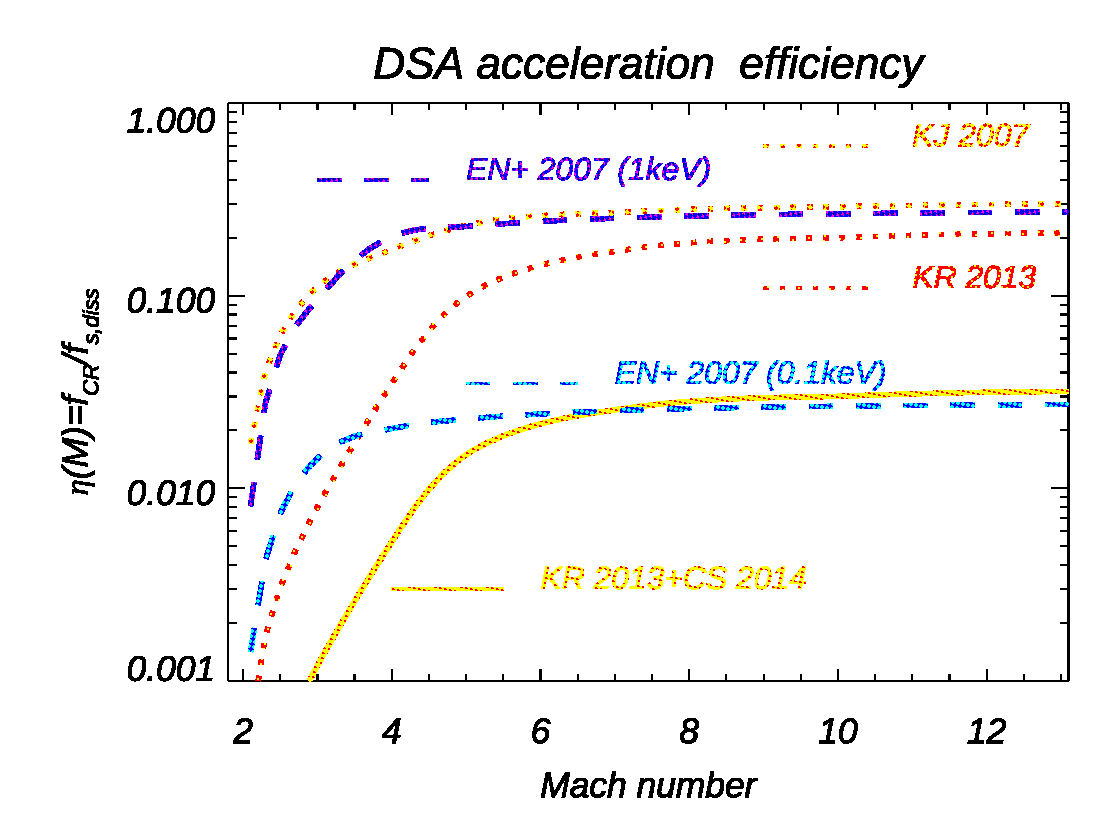} 
    \caption{Different proposed acceleration efficiencies function $\eta(\mathcal{M})$ as a function of shock Mach number, where $\eta(\mathcal{M})=f_{\rm CR}/f_{\rm diss}$ is the ratio between the injected cosmic ray energy flux and the total kinetic energy dissipated by shocks. The various models in different colors are: KJ2007 for \citet{kj07}, KR2013 for \citet{kr13}, EN+2007 for \citet{en07} (including two different gas temperatures) and KR 2013+CS 2014 for an "hybrid model" derived by combining the results of \citet{kr13} and the obliquity-dependent results from \citet{2014ApJ...783...91C}, as in \citet{scienzo16}.} \label{fig:eta}
\end{figure*}

A crucial problem of every cosmological simulation is that the CR acceleration efficiency, $\eta({\mathcal M})=f_{\rm CR}/f_{\rm diss}$ , defined as  the ratio between the injected CR energy flux and the total kinetic energy dissipated by shocks, is unknown and must be guessed based on other works. 
The present state of affairs is summarized in Figure \ref{fig:eta}, which shows several relevant examples of acceleration efficiency functions considered by 
a few cosmological simulations from the  last decade. The KJ2007 model is based on 1-dimensional diffusion-convection simulations of diffusive shock acceleration by \citet{kj07}, while the  KR2013 model is an updated version of the same baseline mode, including the effect of energy dissipation by Alfven waves amplified at the shock, yielding a lower efficiency  \citep{kr13}.  Both models were included in Eulerian simulations \citep[][]{scienzo,scienzo16}, also including the effect of reaccelerated cosmic rays. 
The EN+2007 model is similar to the KJ 2007 model, but assumes a fixed acceleration efficiency (0.3 or 0.03) as a function of the gas temperature (T=1keV or T=0.1keV in the Figure, respectively).  This acceleration model was used in SPH simulations \citep[][]{pf07, 2008MNRAS.385.1211P,pi10}. 

Finally, the KR2013+CS2014 hybrid model is an updated version of the \citet{kr13} model, which includes the effect of a distribution of shock obliquities at cosmic shocks, as discussed in Eulerian grid simulations in \citet{scienzo16} and tested in detail by \citet{wi17}. 
 If  the probability distribution of shock obliquities is random ($P(\theta) \propto \sin(\theta)$), only $\approx 0.3$ of shocks have $\theta \leq 45^{\circ}$(i.e. are quasi-parallel). In this range of shocks,  \citet{2014ApJ...783...91C} have reported that the shock acceleration efficiency is $\approx 0.5$ of the efficiency in \citet{kr13} for the same Mach number range.  Therefore, the combination of these two factors  should roughly mimic the reduced acceleration efficiency of CR protons in quasi-perpendicular shocks in large-scale structures (see also discussion below). 

Clearly, the differences among models are large,  within a factor $\sim 10$ for strong shocks ($\mathcal{M} \geq 6$), but (more worryingly) of order  $\geq 100$ in the regime which is more crucial to cluster merger shocks, $\mathcal{M} \leq  4$.  Most of CRs injected in structure formation shocks should come from these shocks, as well as most of the observed  radio relic emission (e.g. Van Weeren et al., this topical collection), and thus any small difference in the frequency of shocks in this regime, as well as on the assumed acceleration efficiency, will have a dramatic impact in the predicted budget of CRs.  However, it seems now clear that deeper changes in the simplistic view of CR proton acceleration in the ICM must be introduced, in order to cope with observational evidences.

\subsection{Possible solutions to the lack of cosmic ray protons dilemma}

Based on the historic trend of findings in cosmological simulations and after the report of {\sl  Fermi LAT}   non-detections (combined with the observation of radio emission from relativistic electrons in many of the same objects)  
several possible explanations have been explored and tested, all with still inconclusive solutions. \\

A first possibility is to relax the standard assumption that the overall distribution of CR protons remains confined within galaxy clusters for longer than a Hubble time, due to their trapping into the tangled intracluster magnetic field \citep[][]{1996SSRv...75..279V,bbp97}.  A few works have indeed raised the possibility that CR protons may  stream out of the innermost cluster regions, at a velocity $\geq v_a$ (where $v_a$ is the local Alfven velocity) on large scales, which would significantly lower the expected hadronic emission as the signal drops as $\propto n^2$, thereby transforming peaked CR profiles into asymptotically flat ones.
This possibility has been first proposed by \citet[][]{2011A&A...527A..99E} in the case of galaxy clusters, and has been later addressed with more sophisticated methods by other authors 
\citep[][]{2013MNRAS.434.2209W,2014MNRAS.438..124Z,2017MNRAS.465.4800P}. 
At present, the jury is still out on the feasibility of such mechanism in the plasma conditions 
of the ICM, due to the complex and time-dependent interplay between plasma turbulence, MHD waves and CRs, that is expected to be present in every galaxy cluster \citep[][]{2018MNRAS.473.3095W}.  Although theoretically challenging to justify, the $\geq v_a$ streaming of CRs  would in principle alleviate the tension with {\sl  Fermi LAT}  limits and explain the lack of radio emission in radio quiet clusters. However,  this solution cannot account for the lack of hadronic emission in clusters where instead large-scale radio halo emission is observed (see Van Weeren et al., this topical collection), because  at least there CR protons should be present if the emission must be explained away by a secondary electron model.  This makes it overall unlikely that cosmic ray streaming (at least, alone) can explain the lack of detections of hadronic emission reported by {\sl  Fermi LAT}. \\

An other class of possible solutions focuses on the revision of the assumed particle acceleration efficiency of CR protons. 
\citet{va14relics} and \citet{va15relics} made the case of observed "double" radio relics (see Van Weeren et al., this topical collection) which also expose open problems of the "standard" DSA acceleration model applied to the ICM, in the sense that if observed double radio relics were uniquely the product of DSA acceleration, a  significant amount of cosmic ray protons should be present there too. From the modelling of {\it observed} radio relic emission it is indeed possible to guess the shock parameters (including the shock Mach number), as well as the total volume swept in the past by  relics; from the combination of these parameters it is possible to predict the total amount of CR protons that must have been injected (and confined) in the ICM, for different formulations of the DSA scenario. When these estimates are applied to real objects, in several cases the implied injection ratio between electron and proton is unexpectedly high ($\geq 0.1-1$) \citep[][]{va14relics, bj14}.  In other words, the simultaneous radio observation of relativistic electrons and the non detection of 
CR protons via $\gamma$-rays  is explained only if the acceleration from merger shock tend to privilege (or equally accelerate) electrons and protons, which is at variance with expected in the DSA scenario (where the expected efficiency should be $\sim 10^{-5}-10^{-3}$ for typical merger shocks). \\

A solution capable of alleviating part of the problem with radio relics is that part (or most) of the emission in relics comes from shock {\it re-}accelerated electrons, meaning that a pool of "fossil" electrons with a significant energy (albeit not radio emitting) is often present in the volume swept-up by merger shocks \citep[e.g.][]{2005ApJ...627..733M,2013MNRAS.435.1061P}. 
 In this case, the conversion efficiency of shock kinetic energy and radio power gets boosted \citep[e.g.][]{ka12,2013MNRAS.435.1061P} and even relatively weak shocks ($\mathcal{M} \sim 1.3-2.5$) can produce detectable radio emission.  In this scenario, while the CR proton acceleration from merger shocks in the ICM can be considered low enough to never produce a significant level of hadronic emission, the acceleration of CR electrons gets boosted by the presence of a "fossil" population of electrons. 
Therefore,  a key prerequisite for this mechanism to work is the presence of volume filling CR electrons mixed with the thermal ICM.  A large number of radio galaxies (active nowadays or in the past) is surely present in the ICM, and a number of observations reported convincing evidence for a close association between diffuse radio emission an localized sources of seed electrons (e.g. Van Weeren, this topical collection).  Moreover, \citet{2013MNRAS.435.1061P}  have shown that indeed a bath of relic electrons can be present in the outer regions of clusters, with very long cooling times ($\geq t_H$), as a result of shock acceleration in cluster outskirts.  
While this idea can address the presence of relic emission also in shocks with a weak Mach number, in order for this scenario  to cope with the  lack of hadronic emission even in systems with prominent relic emission (suggesting intense shock crossing in the recent past) fossil protons must be absent instead. This may put some tension on the possibility in which 
 fossil cosmic ray electrons come from previous shocks (which would have injected cosmic ray protons as well), and it also requires that the injection of CRs from supernovae and AGN to be overall dominated by cosmic ray electrons, which is non trivial to achieve.  \\

A third possible solution is the incorporation of key "microscopic" details of  the shock electron and proton acceleration, which have been overlooked in the past, at least for problems relevant to galaxy clusters. 
While it is still impossible to fully  include all physical scales which are key to resolve in the simulation of shock acceleration (see Sections above in this chapter), recent cosmological simulations also including magnetic fields have attempted to incorporate recent studies from Particle In Cell (PIC) simulations,  \citep[][]{2014ApJ...783...91C,guo14a}, in which the shock obliquity ($\theta$) is a key parameter to regulate  the acceleration of, both, electrons and protons by cosmic shocks. 

\begin{figure*}
	\centering
    \includegraphics[width=0.49\textwidth]{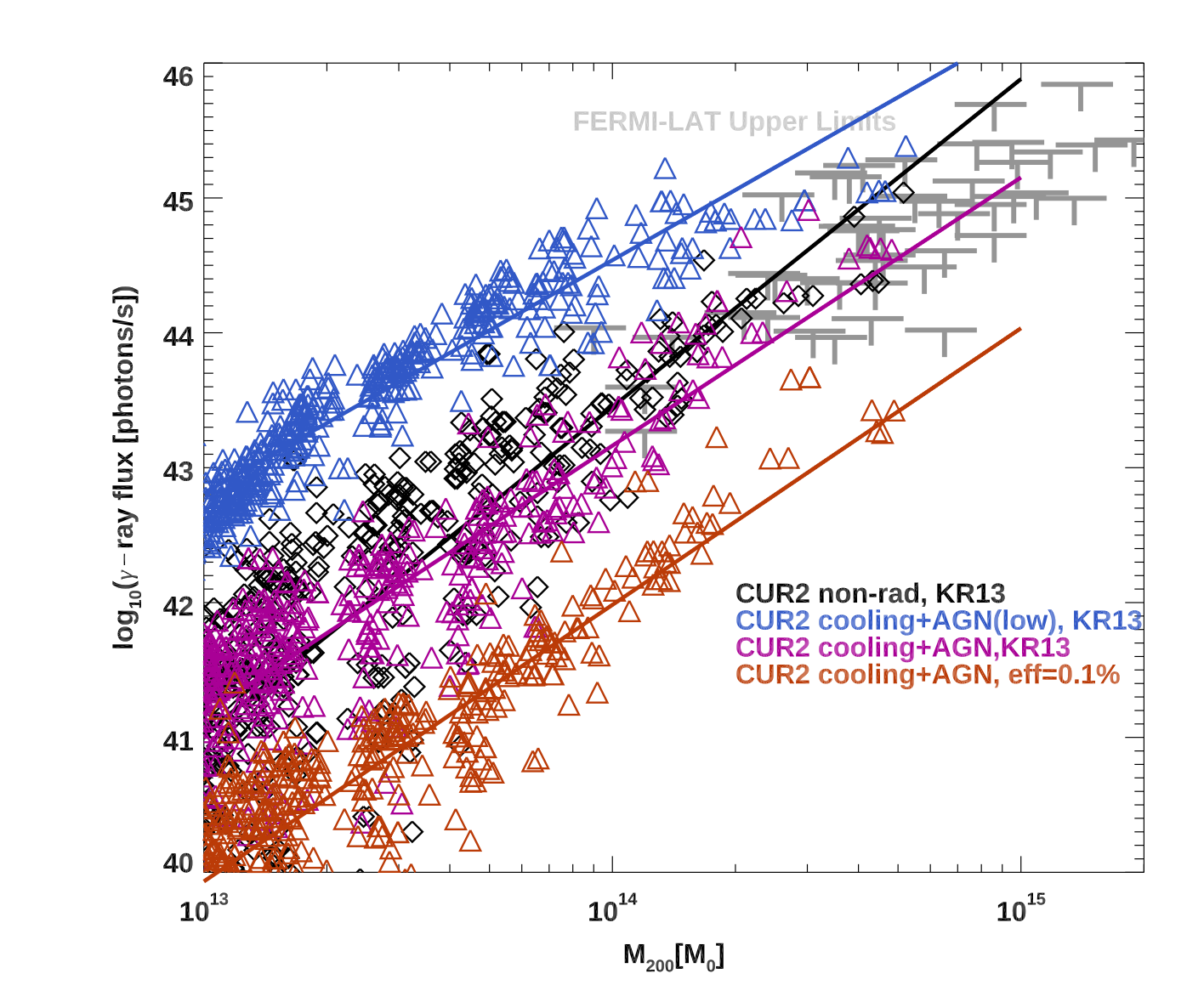}
	\includegraphics[width=0.49\textwidth]{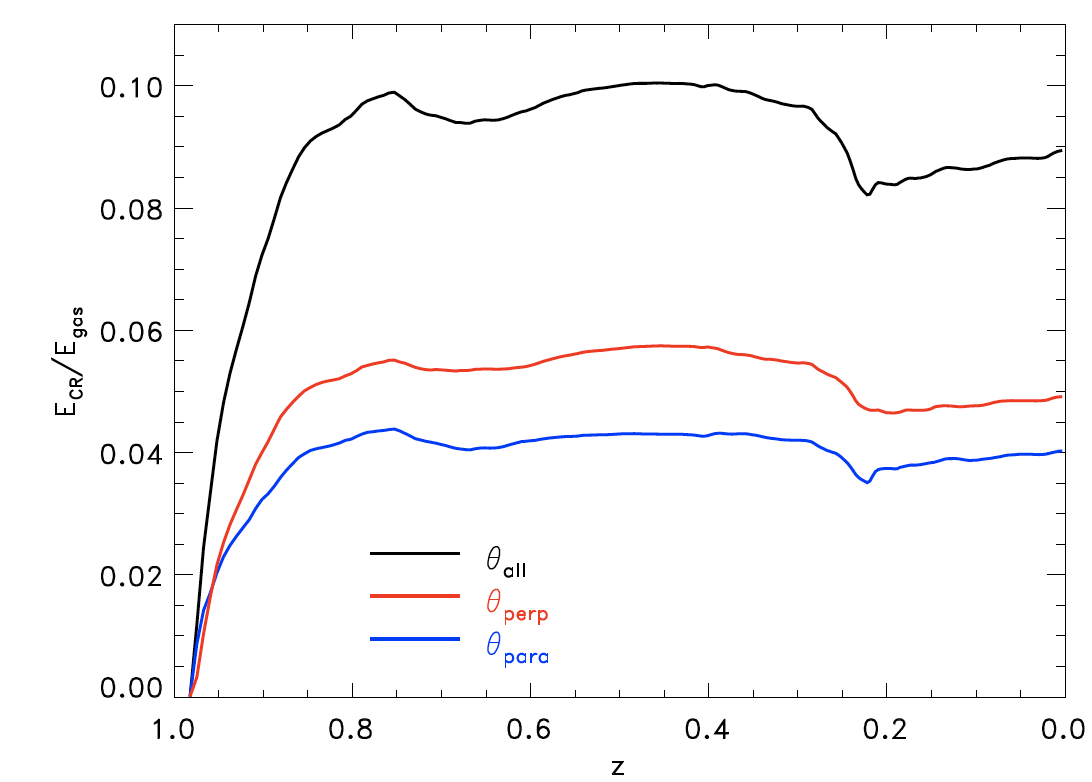}
    \caption{Left panel: hadronic emission for {\enzo}-simulated clusters in \citet{scienzo16} in the 
  0.2-200 GeV energy range. The gray symbols are the upper limits from the {\sl  Fermi LAT} catalog in the same energy range \citep[][]{fermi14}. The different colors refer to resimulations of the same set of galaxy clusters, with  variations of the assumed gas physics (i.e.  non-radiative vs radiative gas physics, AGN feedback) and of the assume cosmic ray acceleration efficiency ("KR13"=acceleration model assumed from \citealt{kr13}; "eff=0.1\%" assumes a fixed $10^{-3}$ acceleration efficiency of cosmic ray protons regardless of the shock Mach number. Taken from \citet{scienzo16}. Right panel:  evolution of the ration between the CR energy and the thermal gas energy for particles accreted onto a  $\sim 10^{15} M_{\odot}$ simulated cluster in \citet{wi17}, for the DSA efficiency model by \citealt{kr13}, or by further limiting the acceleration to quasi-parallel ($\theta_{\rm para}$) or quasi-perpendicular ($\theta_{\rm perp}$) shocks.}\label{fig:ratio1}
\end{figure*}

In the Earth's bow shock and in interplanetary shocks, electrons are {\it directly} observed to be accelerated to relativistic energies preferentially in quasi-perpendicular configurations, i.e. $\theta > 40-50^{\circ}$  \citep[e.g.][]{1999Ap&SS.264..481S,2006GeoRL..3324104O}. Recent PIC simulations of electron acceleration by weak intracluster shocks observed that only quasi-perpendicular shocks can pre-accelerate electrons, via shock-drift-acceleration (SDA), before being injected into DSA \citep{guo14,guo14a}. 
With highly spatially and temporally resolved simulations of forming clusters, \citet[][]{wi16} and \citet{wi17} have tested the above ideas, making the simplifying  approximation that the acceleration of CR electrons is entirely "switched off" for  quasi-parallel shocks ($\theta \leq 50^{\circ}$), while conversely the acceleration of CR protons is entirely "switched off" for quasi-perpendicular shocks ($\theta \leq 50^{\circ}$), based on the results of \citet{guo14} and \citet{2014ApJ...783...91C}.  This introduces an angle-dependent acceleration efficiency in the cosmological run, $\eta({\mathcal M}, \theta)$, which gets  very small for cosmic ray protons during merger shocks  at low redshift, because the ICM magnetic fields is tangled for $\leq \rm ~Mpc$ scales (see e.g. Donnert et al.  this topical collection) and the  distribution of shock obliquities gets close to random, thus  in this regime the majority of shocks is quasi-perpendicular.  As a consequence of this, the energy ratio between CR protons and thermal energy is reduced by a factor $\sim 50\%$ if only quasi-parallel shocks are concerned, and the predicted hadronic $\gamma$-ray emission within $R_{\rm 200} $ gets reduced to $\sim 25-30 \%$ of more standard estimate  (see right panel of Fig.\ref{fig:ratio1}).   This effect is in general not large enough to  fully alleviate the tension with {\sl  Fermi LAT} non detections, yet it suggests  a new level of complexity and realism to the future testing of DSA models.

As already commented in Sec.~\ref{PICshock}, \citet{2018ApJ...856...33K} reported that for quasi-parallel shocks and ICM conditions ($\mathcal{M} \sim 2-3$), when postshock waves are isotropized by plasma processes,  the  Alfv\'{e}nic drift could reduce the injection efficiency of CRs by a factor $\sim 5$ with respect to more standard estimates.

In summary, from the stringent comparison with {\sl  Fermi LAT} non detections of hadronic $\gamma$-ray emission, cosmological simulations have established that only a negligible energy in relativistic CR protons can be present in the intracluster medium (i.e. $\leq 1\%$ of the total gas energy within the virial volume), for typical spectra of accelerated CRs with a power law index $\sim 2.1-2.3$, resulting from the combination of accretion and merger shocks.  This poses challenges possibly to both our understanding of the injection of CR protons across cosmic environment, as well as of the advection and diffusion of CRs during clusters'evolution.  However, the non-thermal components with the steep spectra of the power law indexes $\gsim$  3 can be produced by an ensemble of long-wavelength plasma motions with weak shocks  as it was discussed in section \S\ref{NLensemble}.
The soft components  could contribute more into the nonthermal pressure in the intracluster medium.  Such soft component is observed in the interplanetary medium \citep[][]{2015JPhCS.642a2009F}. Since the nonthermal component is dominated by sub-relativistic ions the sensitive high resolution X-ray spectroscopy may help to constrain it. The departures from the Maxwellian momentum distribution of the Fe ions at the cluster shocks can be potentially constrained with the next generation large effective area spectroscopy instruments aboard the {\sl ATHENA} mission \citep[see e.g.][]{2018SPIE10699E..1GB} and the X-Ray Microcalorimeter-Imaging Spectrometer which is planned for the {\sl Lynx} observatory. \\
While it is preliminary to say if any of the above proposed solutions (or a combination of them) will be able to explain the missing hadronic emission from the ICM, it seems clear that only by jointly  testing the "microscopic" view of particle acceleration at shocks and of the "macroscopic" view enabled by cosmological simulations of forming structures it will be possible to quantitatively test solutions to the challenges posed by radio and  $\gamma$-ray observations. 

\begin{acknowledgements}
 A.M.B.  thanks the staff of ISSI for their generous
hospitality and assistance. The authors thank the referees for the constructive comments. 
A.M.~Bykov and J.A.~Kropotina were
supported by the RSF grant 16-12-10225. Some of the modeling was
performed at the ``Tornado'' subsystem of the St.~Petersburg
Polytechnic University supercomputing center and at the JSCC RAS. F.V. and acknowledges financial support from the European Union's Horizon 2020 program under the ERC Starting Grant "MAGCOW", no. 714196. 
\end{acknowledgements}


\end{document}